\documentclass[sigconf]{acmart}

\usepackage{dirtytalk} % for quotations \say{something}
\usepackage{wrapfig}

\usepackage{booktabs}

\usepackage{subcaption}

\usepackage[bottom]{footmisc}
\usepackage{gensymb}

\definecolor{myRed}{RGB}{255, 45, 85}
\definecolor{myGreen}{RGB}{76, 217, 100}
\definecolor{myBlue}{RGB}{0, 122, 255}
\definecolor{myOrange}{RGB}{255, 136, 0}
\definecolor{darkergreen}{RGB}{37, 155, 57}
\definecolor{myTeal}{RGB}{181, 255, 251}
\definecolor{myLighterGray}{RGB}{213, 217, 224}
\definecolor{myLightGray}{RGB}{177, 177, 177}

\newcommand{\blue}[1]{{\color{black}{#1}}}

\AtBeginDocument{%
  \providecommand\BibTeX{{%
    \normalfont B\kern-0.5em{\scshape i\kern-0.25em b}\kern-0.8em\TeX}}}

\copyrightyear{2023} 
\acmYear{2023} 
\setcopyright{acmlicensed}\acmConference[CHI '23]{Proceedings of the 2023 CHI Conference on Human Factors in Computing Systems}{April 23--28, 2023}{Hamburg, Germany}
\acmBooktitle{Proceedings of the 2023 CHI Conference on Human Factors in Computing Systems (CHI '23), April 23--28, 2023, Hamburg, Germany}
\acmPrice{15.00}
\acmDOI{10.1145/3544548.3580896}
\acmISBN{978-1-4503-9421-5/23/04}

\begin{document}

\title[Stargazer: An Interactive Camera Robot for Capturing How-To Videos]{Stargazer: An Interactive Camera Robot for Capturing How-To Videos Based on Subtle Instructor Cues}



\author{Jiannan Li}
\affiliation{%
  \institution{University of Toronto}
  \city{Toronto}
  \country{Canada}
}

\author{Mauricio Sousa}
\affiliation{%
  \institution{University of Toronto}
  \city{Toronto}
  \country{Canada}
}

\author{Karthik Mahadevan}
\affiliation{%
  \institution{University of Toronto}
  \city{Toronto}
  \country{Canada}
}

\author{Bryan Wang}
\affiliation{%
  \institution{University of Toronto}
  \city{Toronto}
  \country{Canada}
}

\author{Paula Akemi Aoyaui}
\affiliation{%
  \institution{University of Toronto}
  \city{Toronto}
  \country{Canada}
}

\author{Nicole Yu}
\affiliation{%
  \institution{University of Toronto}
  \city{Toronto}
  \country{Canada}
}

\author{Angela Yang}
\affiliation{%
  \institution{University of Toronto}
  \city{Toronto}
  \country{Canada}
}

\author{Ravin Balakrishnan}
\affiliation{%
  \institution{University of Toronto}
  \city{Toronto}
  \country{Canada}
}

\author{Anthony Tang}
\affiliation{%
  \institution{Singapore Management University}
  %\city{Singapore}
  \country{Singapore}
}

\author{Tovi Grossman}
\affiliation{%
  \institution{University of Toronto}
  \city{Toronto}
  \country{Canada}
}

\renewcommand{\shortauthors}{Li, et al.}

\begin{abstract}
% my proposal -- ms
%Live and pre-recorded video tutorials are an effective means for teaching physical skills such as cooking or prototyping electronics. 
%A dedicated cameraperson following an instructor's activities can improve production quality.
%However, instructors who do not have access to a cameraperson's help must work within the constraints of static cameras as they do not . 
%We present Stargazer, a novel approach for assisting with tutorial content creation with a camera robot that dynamically adjusts its viewpoint through changes in position and orientation. 
%Our approach estimates regions of interest and smoothly track them based on sensed instructor actions. Stargazer adjusts its camera behaviors in response to subtle instructor cues, including gestures and speech, allowing the instructor to fluidly integrate camera control commands into instructional activities. 
%Our user study with six instructors, each teaching a distinct skill, showed that participants could create dynamic tutorial videos using Stargazer with a diverse range of camera subjects, framing, and angles, and found their videos effective in communicating their tutorial content.

Live and pre-recorded video tutorials are an effective means for teaching physical skills such as cooking or prototyping electronics.
A dedicated cameraperson following an instructor's activities can improve production quality.
However, instructors who do not have access to a cameraperson's help often have to work within the constraints of static cameras. 
We present Stargazer, a novel approach for assisting with tutorial content creation with a camera robot that autonomously tracks regions of interest based on instructor actions to capture dynamic shots.
Instructors can adjust the camera behaviors of Stargazer with subtle cues, including gestures and speech, allowing them to fluidly integrate camera control commands into instructional activities. 
Our user study with six instructors, each teaching a distinct skill, showed that participants could create dynamic tutorial videos with a diverse range of subjects, camera framing, and camera angle combinations using Stargazer. 
% Participants also found their videos effective in communicating the tutorial content.

\end{abstract}




\begin{teaserfigure}
  \includegraphics[width=\columnwidth]{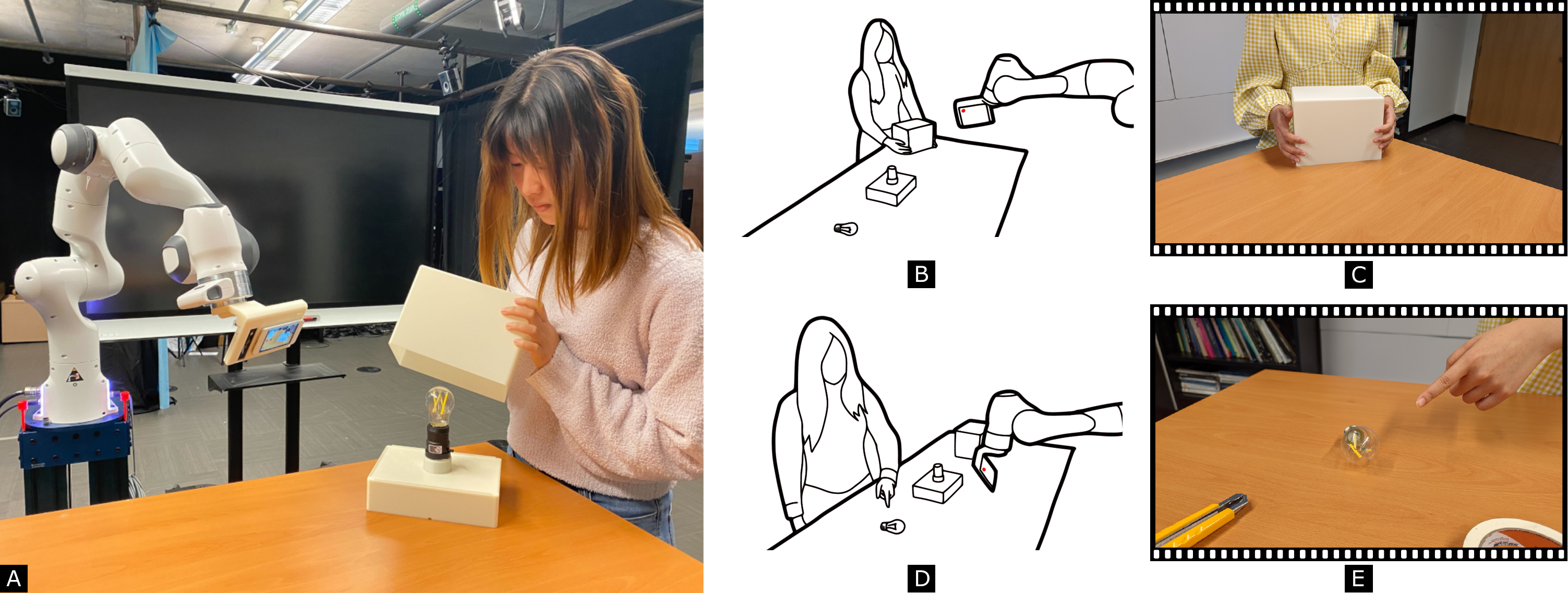}
  \caption{(A) Stargazer is a novel approach for how-to video creation using a camera robot. (B \& C) Stargazer's camera automatically follows the region of interesting, such as the hands of instructor who is holding the cover for a lamp. (D \& E) Stargazer also detects the instructor's signals, such as pointing gestures, and responds by focusing on the light bulb.  }
  \Description{This figure shows three subfigures, where A on the left shows a person lifting the lid of a lamp and a camera robot is filming her actions, B and C shows the person at a corner of the table and the robot films her, D and E shows the person pointing at a light bulb and the robot films the light bulb.}
  \label{fig:teaser}
\end{teaserfigure}

\maketitle

\section{Introduction}
People use how-to videos, both live and pre-recorded, to learn new physical skills~\cite{mayer2020five}, ranging from repairing a broken keyboard to learning digital fabrication~\cite{peek2021making}. 
% With the push towards remote learning (accelerated by the global pandemic), instructors have been pushed to experiment with remotely teaching hands-on skills such as creating breadboard circuits and .
In most how-to videos for physical skills, the instructor demonstrates step-by-step how to complete the task~\cite{bjorn2018developments}. 
%This is often complicated by the limitations of static cameras, since steps involving physical manipulations could be missed due to occlusion or poor perspective. 
As these steps may involve activities at varying locations in varying levels of detail, a single, fixed camera often cannot record every step with the desired clarity~\cite{mok2017critiquing}. 
This necessitates frequent changes to camera parameters, including viewpoints, angles, and zoom levels.

Professional video productions, such as cooking and home improvement shows, employ several dedicated camera operators who actively re-position cameras and adjust their parameters in response to the instructor's actions. 
However, such resources are not available to most instructors; 
%(particularly amateurs)
instead, these rely on one or more pre-configured fixed cameras.
%instead they have to play the role of the demonstrator and camera operator. Consequently, instructors are often constrained to static camera setups. 
Although fixed camera setups can be re-configured during recording, instructors need to stop what they are demonstrating (e.g., chopping vegetables) to manipulate the camera.
%This makes filming bi-manual actions impossible and any interference must be removed after filming (during post-processing).
This disrupts the demonstration, increasing the instructor's workload. 
It also requires more post-processing to combine clips filmed with different camera setups.  

Both filmmakers and researchers have explored the idea of camera-manipulating robots as an alternative to human operators~\cite{mrmc2022bolt,joubert2016towards}. 
Recent camera robots (predominantly drones) can autonomously track moving subjects~\cite{nageli2017realtime,huang2018act,bonatti2020autonomous}. 
However, it remains a challenge for the user being filmed to control the camera robots' behaviors while performing other activities, such as demonstrating a physical process.
Conventional interfaces for robot control employ joysticks~\cite{sheridan1986human}, gestures~\cite{sauppe2014robot}, and speech~\cite{forbes2015robot}---all of which require dedicated input actions---that disrupt instruction delivery.
\blue{
If not edited out, such disruptions might split audience's attention and hinder learning~\cite{cerbin2018improving}, but post-processing adds to instructors' efforts.}
Recent user interface research has explored triggering on-screen visual effects through presenters' gestures and speech~\cite{saquib2019interactive,hall2022augmented,liao2022realitytalk} that are part of the presentations. 
Our approach in this work is to: (1) identify the kinds of camera shots that how-to videos use and (2) direct camera operations in a non-disruptive manner by relying on the communicative signals that instructors already use to address their \textit{audience} during demonstrations. 
For instance, an instructor may point to a part of an object to emphasize it, use speech to guide the audience's attention, or wave to introduce themselves.

We present Stargazer, a novel approach that uses a camera robot to capture dynamic how-to videos for tabletop-scale physical tasks (Figure~\ref{fig:teaser}A). 
Instructors can use subtle signals to control the high-level behavior of the camera, such as the current region of interest and camera framing, in real-time.
Stargazer's camera autonomously tracks and captures the region of interest (Figure~\ref{fig:teaser}B-E).
%, following cinematography principles and instructor specified camera framing and angles.   
% autonomously tracks areas of interest while reacting to the instructor in real time. 
Note that our goal is not to build a fully automated robot that understands every nuance of human communicative behavior; instead, we aim to explore a vocabulary of controls that can support the necessary camera work needed to create informative tutorial videos and can be fluidly blended into normal instruction activities.
%Similar to a human camera operator, Stargazer gleans these signals and adjust its behavior accordingly. 
%Building a fully autonomous robotic camera that recognizes and reacts to all nuances of human communicative behavior is out of scope. 
%Instead, we use Stargazer to explore the design space of \textit{control signals} that an instructor can use to orchestrate the creation of video content which are disguised as part of the demonstration of a physical task.

% While these cues are primary meant to address the \textit{audience}, they are also picked up by human camerapersons to determine camera motion. 

Based on an analysis of 50 how-to videos under three popular categories, we identified three common types of shots and three types of camera parameter control that the robot should support. 
Instructors control shot types and camera parameters through gestures and speech that they commonly use to guide the attention of their audience. 
%The three types of shots are characterized by the subject in focus: \textit{instructor shots} which focus on the instructor's face, \textit{action shots} which emphasize the instructor's hands, and \textit{object shots} which focus on objects in the workspace. % The three communicate cues include gaze direction, deictic hand gestures, and directional utterances. 
Stargazer locates the subject of interest in each type of shot and smoothly tracks them. 
%The camera parameter controls include camera framing, angles, and movements.
 
%Using these cues, the instructor can guide the audience as well as the robot to make adjustments to framing or transition between shots.
% It recognizes the five cues  

We invited six participants, including two professional filmmakers, to capture how-to videos with Stargazer for a physical task of their expertise.
All participants were able to create videos without needing any dedicated controls other than the cues afforded by Stargazer and were satisfied with the quality of the video produced.
Based on participants' think-aloud comments on the produced videos and post-study interview data, we identify opportunities and challenges for collaborative video capture with robots.

%Participants were overall satisfied with the quality of the outcomes, and suggested \lightGray{some excellent ideas}.  

\section{Related Work}
%Our work extends the long  
Our research builds on prior work on how-to videos, camera robots, and human cues for robot control.
Our interaction design is inspired by recent progress in performance-driven interactive presentations.
Below we review related work.

\subsection{How-To Videos}
How-to videos are commonplace on online video and livestreaming websites and provide convenient ways to learn physical tasks~\cite{mayer2020five,bjorn2018developments,fraser2019sharing}.
Research in human-computer interaction (HCI), cognitive science, and education technologies has studied how attributes of how-to videos, such as perspectives~\cite{huang2022immersive}, pace~\cite{tuncer2020on}, and instructor identities~\cite{hoogerheide2018model}, could impact learning.  
User interface research has specifically investigated novel interfaces for exploring~\cite{chang2018recipescape}, navigating~\cite{chang2021rubyslipper,zhao2022rewind}, annotating~\cite{kim2014crowdsourcing}, and editing~\cite{chi2013democut} how-to videos. 
Our work builds on these previous works on how-to videos but aims to facilitate filming these videos instead of their post-processing or consumption.

\subsection{Interactive Presentations}
%People naturally leverages a 
Researchers have explored how the performance of the presenter, including speech~\cite{liao2022realitytalk}, gesture~\cite{saquib2019interactive,hall2022augmented}, and sketching~\cite{perlin2018chalktalk,suzuki2020realitysketch}, can drive real-time on-screen graphics to make live or recorded presentations more informative and engaging.
A particularly elegant aspect of these interactive presentation interfaces is that they allow presenters to merge graphics effect controls into content delivery.
RealityTalk~\cite{liao2022realitytalk} leveraged keywords in verbal presentations to trigger visual augmentations.
Saquib et al.~\cite{saquib2019interactive} enabled presenters to directly manipulate graphical elements and effects with gestures and changes in postures. 
Hall et al.~\cite{hall2022augmented} also studied the use of gestures, focusing on modifying data visualizations.
Our user interactions with Stargazer are inspired by how prior work fluidly fits effect controls into presenter performance by leveraging gestures and speech.  
Instead of controlling digital artifacts captured by a fixed camera, Stargazer focuses on controlling a dynamic camera with instructor actions.

\subsection{Human Cues for Human-Robot Interaction}
Prior research has examined coordination mechanisms that human-human teams use for collaborative physical tasks.
%involving \textit{joint action}~\cite{mutlu2013coordination} (e.g. assembly). 
These tasks demand joint attention---directing a partner's attention to an object or area of interest.
%such as assembling a transmission. Although demonstrating physical skills in how-to videos only requires the human to perform the task while the robot captures it, the coordination mechanism of \textit{joint attention}---which involves directing their partners' attention to an object or area of interest, is especially pertinent for the design of Stargazer as directing the robot's attention is reflected in what the audience can see in the final video.
Human-robot interaction researchers have proposed several methods to aid joint attention. 
\blue{
A direct approach can be having users annotate the robots' sensor input (e.g., camera streams)~\cite{yamamoto2022photographic,senft2021situated}, but this usually requires an additional device.
}
Tracking the human's gaze enables anticipatory robot planning and execution of actions~\cite{huang2016anticipatory,admoni2016predicting, aronson2020eye, newman2020examining, aronson2021inferring}, or can be used to prompt robot takeover when it detects user hesitation~\cite{sakita2004flexible}. 
%\textit{Anticipatory assistance} can support shared control of robot arms~\cite{}. 
A gesture is another mechanism to direct attention by humans and robots~\cite{sauppe2014robot}, including pointing, presenting, and exhibiting objects while collaborating on physical tasks~\cite{gleeson2013gestures}. Humans can even use language to direct a robot's attention~\cite{tellex2020robots}, which has been utilized for hands-free robot programming~\cite{forbes2015robot} and commanding a robot during pick-and-place tasks~\cite{matuszek2014learning}.
We note that prior research has typically used either explicit communication mechanisms (i.e., pointing, commands, etc.) or more implicit communication mechanisms (i.e., eye gaze) to manipulate the robot.
Stargazer's approach employs both explicit and implicit communication acts. Still, our approach considers these acts part of the performative dialogue between the presenter and the \textit{audience} rather than between the presenter and the robot.

% Of particular interest to our work is interpreting spatial relationships between objects; researchers have explored frameworks that enable users to provide more abstract commands such as ``place the apple in the bowl''~\cite{alomari2017natural, paul2016efficient}. In Stargazer, spatial relationships are helpful for instructors to frame shots (e.g., overhead versus close-ups).

% multimodal (Huang)
%Prior work has also combined modalities to aid joint attention. For instance, gaze can reinforcement pointing gesture~\cite{sauppe2014robot} or spoken dialogue~\cite{staudte2011investigating}. 

% Directing attention
%Here we focus on prior work that facilitates the understanding of user intent by the robot which is unidirectional.

%Here we focus on prior work that facilitates unidirectional communication, since Stargazer infers user intent and proactively acts (or reacts).

%and \textit{action coordination}---involving partners performing complementary actions. Here we focus on prior work that facilitates unidirectional communication, as the goal is for the robot to infer the user's intent and proactively act or react.

% Action coordination

% Write this somewhere: Though the user never directly manipulates stargazer, the techniques above are used to direct its attention and coordinate actions with the eventual goal of creating an instruction video.

\subsection{Camera Robots}
\blue{Research in remote assistance and telepresence mostly studied camera robots that were manually operated with GUIs by remote users (e.g.,~\cite{gurevich2012teleadvisor,tsui2011exploring,villanueva2021robotar,li2022asteroids}).}
Recent remote assistance camera robots, such as RemoteCoDe~\cite{sakashita2022remotecode}, can be controlled implicitly by remote helpers' face orientation.
Another line of work has explored camera robots that react to signals from the people being filmed. 
Autonomous camera drones can track the human subject and adjust their flight plans based on the subject's positions~\cite{nageli2017realtime}, postures~\cite{huang2018act}, and actions~\cite{bonatti2020autonomous}, often informed by cinematography principles~\cite{joubert2016towards} or learned preferences~\cite{bonatti2020autonomous}.  
While not filming humans, Rakita et al.~\cite{rakita2018autonomous} designed a method for a robot arm to dynamically capture another arm being teleoperated to assist teleoperation.
Although responsive to human signals, these robots mainly do not allow the user being filmed to actively initiate significant behavior changes in the robots, such as changing the subject or camera framing.
Stargazer aims to fill this gap by providing instructors with more fine-grained real-time camera behavior controls for filming how-to videos. 

%\subsection{Subtle Interactions}

%\blue{I think we are inspired by the stuff about subtle commands (or hiding actions) -- something by Fraser and maybe Tovi a few years ago? Not sure if it needs its own section, but I think it is worth calling that kind of stuff out explicitly. 

%Depending on how things are built right now, these are either (a) still "explicit commands" from the perspective of computer, or (b) are part of a weighted optimizing function (so, less explicit); BUT, from the perspective of the audience, they are intended to blend into the performance of the instructor
%}
\section{Characterizing Camera Use in How-To Videos} \label{sec:video-analysis}

The design of Stargazer's behaviors and interactions was informed by studying the content and visual language of existing how-to videos. 
We studied 50 videos from three categories of physical skill instruction with the aim of learning: (1) What are the common types of shots in these videos? and (2) How are framing, angles, and movement used in how-to videos by creators?
%shot typesthe subjects captured in individual shots? (2) What are the common cinematographic choices, including framing, angles, and movements, that how-to video creators make to deliver their instructions? 
% Our analysis focused on the type of shots seen in these videos and cues that instructors employ to direct the attention of their audience, and effectively, the viewpoint of the camera.

\begin{figure*}[t]
  \centering
  \includegraphics[width=\textwidth]{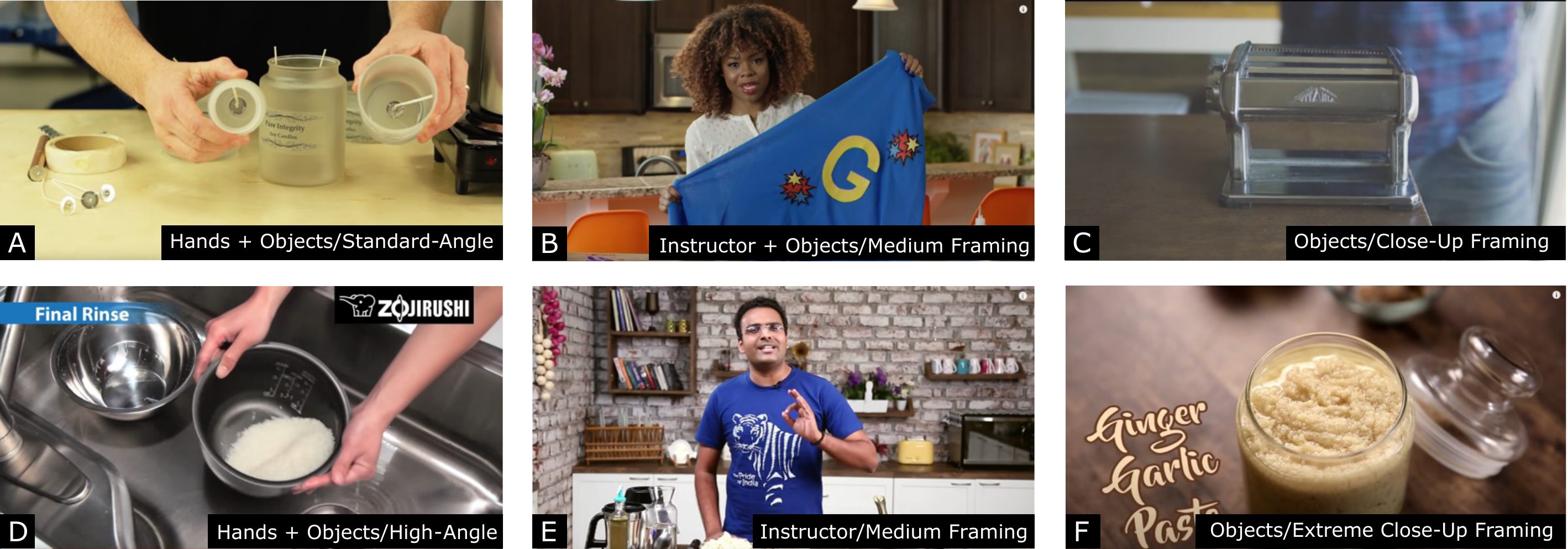}
  \caption{Example frames from the video analysis. (A) a \textit{hands+objects} shot from the standard camera angle (B) an \textit{instructor+objects} shot with medium framing (C) an \textit{objects} shot with close-up framing (D) a \textit{hands+objects} shot with from a high camera angle (E) an \textit{instructor} shot with medium framing (F) an \textit{objects} shot with extreme close-up framing}
  \Description{This figure shows six example frames from the video analysis. (A) a hands+objects shot from the standard camera angle (B) an instructor+objects shot with medium framing (C) an object shot with close-up framing (D) a hands+objects shot with from a high camera angle (E) an instructor shot with medium framing (F) an object shot with extreme close-up framing}
  \label{fig:sample-frames}
\end{figure*}

\subsection{Video Corpus}
%Our goals when sampling instructional videos to study is to collect high-quality tutorials for popular physical tasks. 

Our video corpus contains 50 instructional videos, with 15 on Computers and Electronics, 18 on Food and Entertainment, and 17 on Hobbies and Crafts. 
The average video length is 5 minutes 22 seconds.
The average number of views was 2,176,470 (min: 20,824; max: 37,067,768) as of July 2022.
We constructed our corpus based on physical tasks on  WikiHow~\footnote{https://www.wikihow.com/Main-Page}, a well-known database with thousands of user-produced how-to articles on a wide range of daily tasks, based on strategies used in prior work~\cite{miech2019howto100m}.

% in an open collaboration model -- content is written and edited by individual collaborators and, in some cases, also reviewed by specialists in their fields. 
%WikiHow offers content in multiple languages, however, for this work, we focused primarily on English content. 
To identify classes of physical tabletop-scale tasks, we followed ``Popular Categories'' on the website, and found three categories: ``Computer and Electronics'', ``Hobbies and Crafts'', and ``Food and Entertainment''. 
We then randomly surveyed each category to gather 50 how-to articles and used the article titles to search for videos online on YouTube\footnote{https://www.youtube.com/} using incognito mode.
% to sidestep the recommendation algorithm, which would have influenced search results.
We then selected videos based on the following criteria: 
1) Responds to the how-to question collected from WikiHow;
2) Recorded in the English language;
3) At most 20 minutes long;
4) Accumulated over twenty thousand views;  
We used number of views as a proxy to ensure reasonable quality for the sampled videos. 
\blue{
A recent survey of instructional and informational video viewers found that less than 10\% of their respondents preferred videos longer than 20 minutes~\cite{knott2022video}.
Therefore, we focused on videos at most 20 minutes long to allow our team to analyze a sufficiently diverse dataset within our resource constraints.
}
%Note that this video corpus aims to capture common practices but not necessarily the best practices for making instructional videos for the surveyed topics. 

\subsection{Analysis}

We coded the sampled videos at a shot level (884 shots in total, average of 17.68 shots each), focusing on what content was being captured in each shot and how they were communicated through distinct visual language (i.e., angles and framing). 
% \blue{Do we still want to talk about proficiency?}
We created a bottom-up coding scheme with two major categories: \textit{subject in the shot} and \textit{camera setup} (i.e., angles, framing, and movements). 
% Proficiency relates to how professional the video looks, and we found videos to be amateur, semi-pro or professional according to the level of sophistication in cinematic quality (i.e. lighting set up, camera movement). 
% \textit{Purpose} reveals what the shot is meant to convey to the audience (e.g. show materials, explain rationale).
\textit{Subject in the shot} covers what is being shown (e.g. the host, objects). 
Our coding scheme for \textit{Camera setup} was informed by common cinematography practices~\cite{brown2016cinematography}.  
This category relates to the visual language, including camera angle (e.g., high angle, over-the-shoulder), camera framing (e.g., medium, close-up), and camera movements (e.g. pan, truck, orbit).
Two researchers created and tested the coding scheme by analyzing five videos together and then worked separately on the remaining videos.
%\blue{There is missing detail. Is the unit of analysis "per shot"? How many shots were coded in total? How many shots were there across all videos? What was the average? Did this vary between categories? We don't need all of that, but these are questions that are raised, because it's not clear what is happening with the analysis.

%Also: in a given shot, is there movement? Or, are these fixed shots? Was this coded for?}
%We coded a total of 50 videos from 3 categories,  To ensure the diversity of the videos, we sampled a list of common “how-to” questions from WikiHow and identified instructional YouTube videos that addressed theses questions. 

%Our coding scheme focuses on the content and visual language of creating how-to videos involving physical processes. 

\subsection{Subjects of Shots} \label{sec:purposes-subjects}

%\blue{

%Maybe the ordering in the Table should be different:

%Instructor, Instructor + Object, Hands + Object, Object, Other
%
%[wider --> focused]

%}
%\begin{table*}[!t]
%  \caption{Count of camera shots broken down by Purpose x Subject in the how-to video corpus}
%  \label{tab:purposes-subjects}
%  \begin{tabular}{lccccc}
%    \toprule
%    Purpose & Hands + Object & Instructor & Instructor + Object & Object & Other\\
%    \midrule
%    Introduction & 12 & 21 & 11 & 14 & 13 \\
%    Explain rationale & 14 & 39 & 11 & 14 & 10\\
%    Show materials & 45 & 0 & 6 & 26 & 0\\
%    Show process &  334 & 14 & 55 & 63 & 0\\
%    Show outcome & 42 & 0 & 13 & 22 & 0\\
%    Wrap up & 10 & 7 & 4 & 4 & 11\\
%    Other & 16 & 9 & 5 & 13 & 26 \\
%    \bottomrule
%  \end{tabular}
%\end{table*}
%We identified the purposes of individual shots as well as recurring patterns when these shots are created. Overall, we found that the sampled how-to-videos follow a similar linear structure: introduction -> show materials ->  show process ->  explain rationale -> show outcome -> wrap up. 
%This structure is applicable to videos in all three categories. As expected, most of the time spent in how-to videos is dedicated to describing the physical process step-by-step, taking more than half of a video’s duration on average (56.12\% across all categories).
%\blue{What is "show process" -- is this the bulk of the video? What is "explain rationale" -- not sure what that means.}

We observed that how-to video shots focus on a small set of subjects---the instructor, the actions of the instructor, and the objects used in demonstrating the task. 
The instructor's actions primarily manifest in the form of manipulating tools and other artifacts with their hands. 
How-to video shots often feature both the instructor’s hands and the objects being manipulated (\textit{hands+objects}, 53.51\% of total video length, Figure~\ref{fig:sample-frames} A and D). 
%``Show process'' shots often feature both the instructor’s hands and the objects being manipulated (\textit{hands+objects}, 71.67\% of all ``show process'' shots) or objects-only (\textit{objects}, 13.52\% of all ``show process'' shots). 
%\blue{Careful of significant figures. Also, what makes up the remainder then?}
%We found similar patterns for ``show materials'' and ``show outcomes''. 
%\blue{and, what is it?}
%where the most popular choices of subjects were showing both object and hands, followed by objects only. 
% We further group these three types of shots as \textit{Demonstration Shots}.
% Some videos, but not all, also show comparison between objects and animated transitions with text descriptions.
The second category of shots focuses more on the instructor verbally ``lecturing'' to the audience and thus primarily shows the instructor talking and presenting objects to them (\textit{instructor}, 10.18\% of total video length and \textit{instructor+objects}, 11.88\% of total video length; see Figure~\ref{fig:sample-frames} B and E).
% Thus we group these two types of shots as \textit{Lecturing Shots}.
Finally, 17.65\% of the total video length analyzed includes shots featuring objects alone (\textit{objects}; Figure~\ref{fig:sample-frames} C and F).
These shots often show the materials or tools used in the task or the outcome of the process. 

%Note that tutorial video creators use a wide variety of other shots in their work, often out of nuanced artistic choices.
%Our analysis focus on more general types of shots.
%The boundaries between these categories are also not rigid.

\subsection{Visual Language}
While the visual language in film-making encompasses a myriad of elements, our analysis focused on those relevant to the positioning of cameras, i.e., camera framing, angle, and movements.

\subsubsection{Camera Framing}
We found that shots showing each type of subject mainly use two types of framing.
%, one that frames both the subject and the context and one that just focus on the subject itself. 
Shots showing \textit{hands+objects} or \textit{objects} alone predominantly uses close-up framing (79.92\% for \textit{hands+objects}, 64.74\% for \textit{objects} alone, Figure~\ref{fig:sample-frames}A, C, D), followed by extreme close-up (15.86\% for \textit{hands+objects}, 28.21\% for \textit{objects} alone, Figure~\ref{fig:sample-frames} F).
Shots focusing on the instructor are filmed more with medium framing (66.67\% for the instructor, 87.62\% for \textit{instructor+objects}, Figure~\ref{fig:sample-frames}B and E) than with close-up (33.33\% for \textit{instructor}, 10.48\% for \textit{instructor+objects}). 

\subsubsection{Camera Angles}
When coding camera angles, we used common camera-angle types while adapting them to fit the context of how-to videos. Following cinematography conventions, we labeled shots looking down at the subject (hands/instructor/objects) as \textit{high-angle} and those looking up at the subject as \textit{low-angle}.
Overhead shots were also included in the high-angle category.
Shots from the same level as the subject, or slightly above it, were labeled as \textit{standard}.
We found that most shots focusing on the instructor were made with a standard angle (98.89\% for \textit{instructor} and 82.86\% for \textit{instructor+objects}, Figure~\ref{fig:sample-frames}B and E). 
In contrast, high-angle (22.42\%), point-of-view (34.02\%), and standard angles (29.57\%) are all commonly seen in shots focusing on \textit{hands+objects} and \textit{objects}.  
Figure~\ref{fig:sample-frames}A and D show examples of standard-angle and high-angle shots for \textit{hands+objects}.

\subsubsection{Camera Movements}
The majority (70.59\%) of the shots from our sampled videos were shot with static cameras. 
About 15\% of the shots had shaky, abrupt camera movements, which were most detrimental to their visual quality. 
Only a small proportion (8.13\%) of shots applied stable camera movements, such as pan (rotating the camera around its own vertical axis), truck (moving the camera along a horizontal axis perpendicular to the direction of the lens), orbit (moving in an arc while focusing on the same subject), and dolly (moving towards or away from the subject)~\cite{brown2016cinematography}. 
This is likely because most instructors do not have access to a dedicated cameraperson.

\subsubsection{Summary} Our exploration of camera use in how-to videos helped us to understand the space of camera shots that Stargazer would need to support---what they focus on (the subjects), as well as the range of parameters that would need to be modifiable/used by producers. We next describe the interaction design of Stargazer, which makes use of this vocabulary of shot types, and discuss how Stargazer gives a producer the ability to transition between different types of shots \textit{during} a shoot without needing to break from the instruction or pre-defining camera paths.

\section{Stargazer Interaction Design}

\begin{figure*}[t!]
  \centering
  \includegraphics[width=0.8\linewidth]{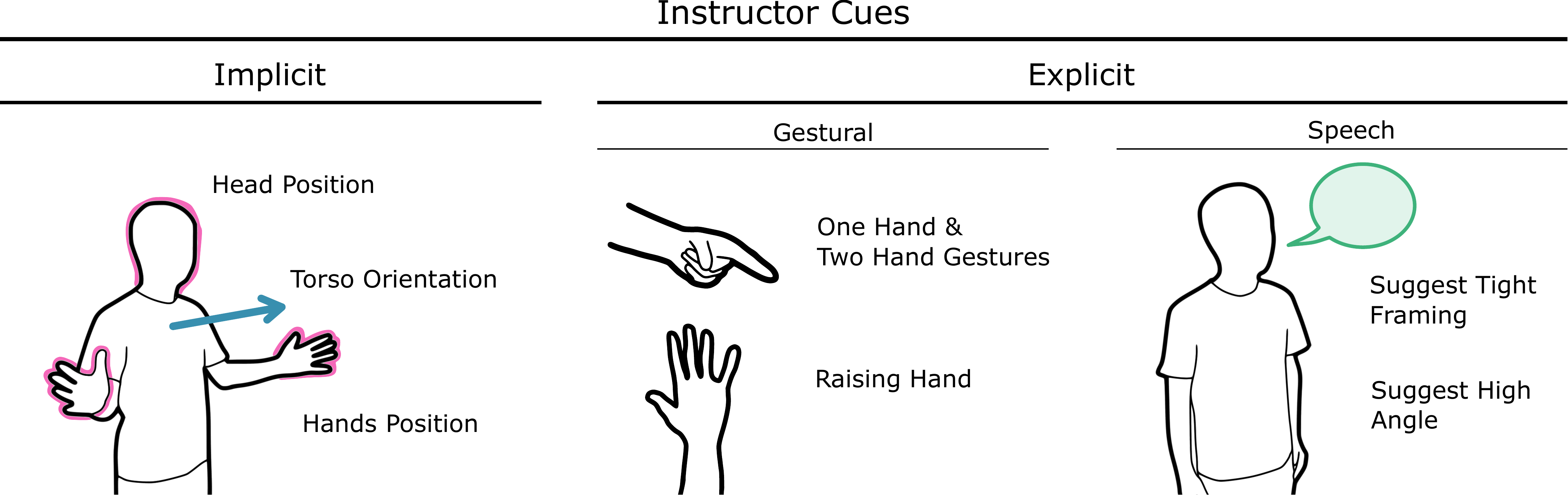}
  \caption{Instructor cues that influence Stargazer behaviors.}
  \Description{This figure shows instructor cues that influence the behaviors of Stargazer, including head position, torso orientation, hands' positions, pointing, raising a hand, and speech}
  \label{fig:interaction-vocabulary}
\end{figure*}

Stargazer helps instructors produce ``one-take'' how-to videos with a single camera on a highly articulated, 7-degree-of-freedom robot arm.
The system tracks subjects and adjusts camera framing and angle dynamically in response to the instructor's actions and speech.
%While Stargazer does not replace a human videographer, Stargazer responds to both subtle cues supplied by the instructor.
%For instance, in a model assembly how-to video, the instructor can signal Stargazer to frame them by a waving gesture they perform to greet the audience.
For instance, the instructor can have Stargazer adjust its view to look at each of the tools they will be using during the tutorial---pointing at the tools one after another to have Stargazer pan to show each of them.
Finally, they can say to the audience, ``Now if you look at how I put A into B from the top'', Stargazer will respond by framing their action with the items with a high angle, giving the audience a better view.

%Stargazer's interaction design is based on the idea an ``intuitive camera operator'' that understands instruction dynamically. 
Our goal in designing the interaction vocabulary was to identify signals that are smooth, subtle, and could conceivably be a part of the instructor's dialogue/rapport with the video audience.
Our design avoids the need for the instructor to communicate separately to the camera robot independent from the audience; instead, the signals are subtly integrated into the tutorial presentation.
%designed to be interpretable by both the audience (i.e., to signal attention), and by the camera (e.g., to signal a camera framing/movement/angle change).

\subsection{Capabilities and Needs}
Our analysis of how-to videos showed a wide range of video shots and capabilities would be needed by instructors.
This provides the building blocks of Stargazer's functionality.
We characterize these based on the vocabulary and coding scheme developed in Section~\ref{sec:video-analysis}.

\begin{itemize}
    %\begin{description}
        \item \textit{Subject:} The camera needs to be able to capture the subjects common in how-to videos.
        %i.e. \textit{instructor}, \textit{hands}, and \textit{objects}.
            \begin{description}
                \item{\textbf{Instructor shots}} primarily frame the instructor for directly addressing the audience---sometimes while presenting objects. (\textit{instructor} and \textit{instructor + objects} from Section~\ref{sec:video-analysis}.)
                \item{\textbf{Action shots}} capture the instructor's hands manipulating various tools and objects when performing the demonstrated physical processes. (\textit{hands + objects} from Section~\ref{sec:video-analysis}.)
                \item{\textbf{Object shots}} focus on one or more objects that the instructor points to (\textit{objects} from Section~\ref{sec:video-analysis}.)
            \end{description}
        \item \textit{Following:} Because the subject may be moving, the robot should automatically locate and follow the subject, maintaining this view within the constraints of the current shot type as well as distance.
        \item \textit{Camera Framing/Angles:} Instructors should be able to select and transition between normal and tight (zoomed-in) framing, as well as standard and high angles.
        \item \textit{Camera Movements:} Instructors should also be able to initiate common cinematographic camera movements (e.g. orbit, truck).
        \item \textit{Feedback:} As with other agents and artifacts that act autonomously~\cite{sheridan1986human,ju2015design}, the robot needs to provide feedback to the instructor about its current state.
    %\end{description}
\end{itemize}

\subsection{Interaction Vocabulary}

Stargazer responds to instructor signals selected to be subtle (Figure~\ref{fig:interaction-vocabulary}).
Our goal is to define a vocabulary that ideally fits into the natural flow of the content delivery, without being jarring to the audience and the instructor.
The vast majority of the time, Stargazer is understood to be operating autonomously in a reasonable way (i.e., tracking and following the subject with a given framing), and smoothly transitioning between shots to increase legibility.
The default camera behavior follows the implicit bodily movements of the instructor and the objects as they are being manipulated.
Shifts in the camera position and orientation are driven by signals from the instructors---the robot constantly monitors the instructor's hands, head, and body postures looking for these signals.
Gesture and speech signals from the instructor---ostensibly to the audience---signal to Stargazer changes in high-level camera behaviors, including the subject of shots, camera framing, and camera angles.
Our designs select cues that find a balance between subtlety and the chance of false activation.

\begin{figure}[!t]
  \centering
  \includegraphics[width=\columnwidth]{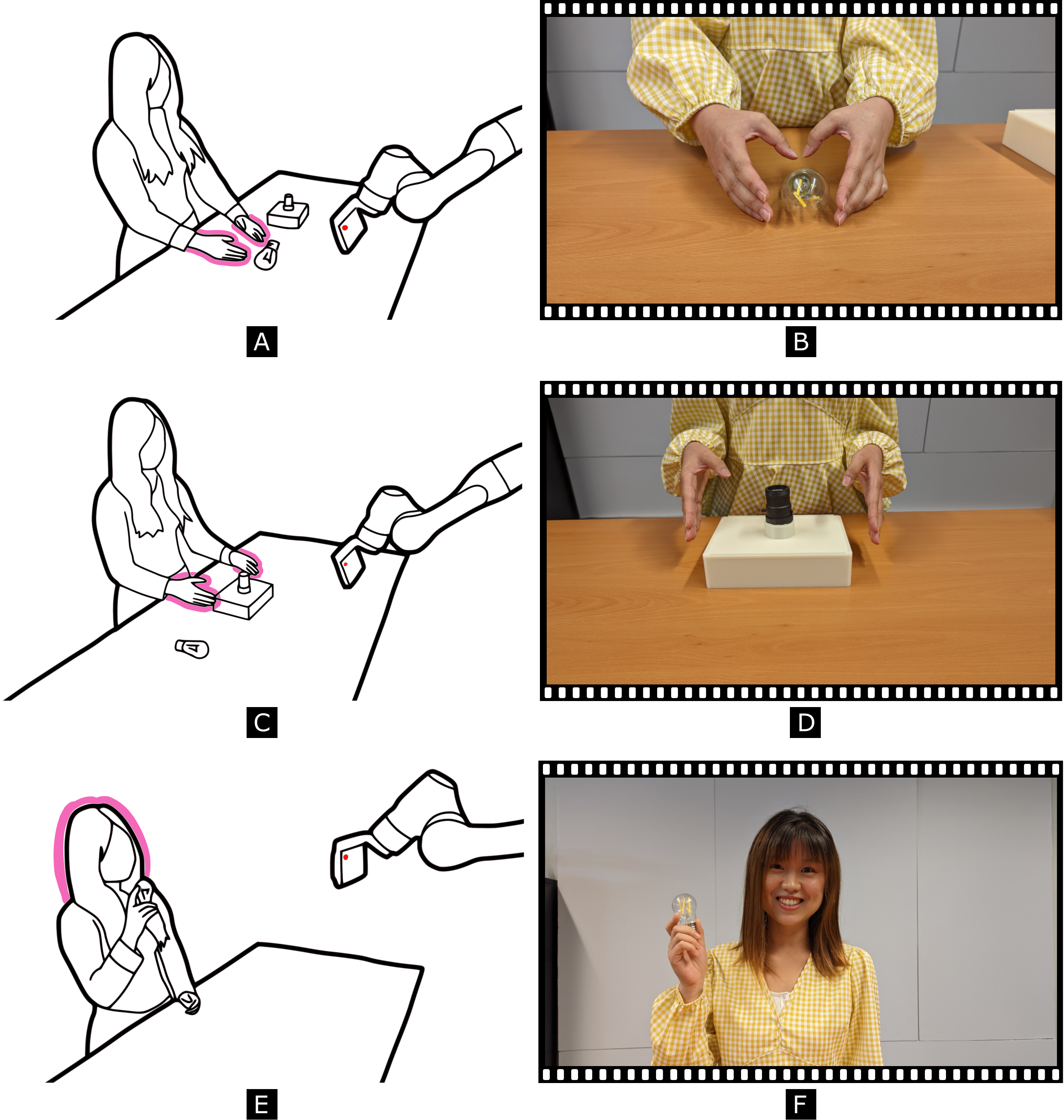}
  \caption{Body movement cues that Stargazer reacts to. (A and C) In action shots, Stargazer follows the hands of the instructor and adjusts its distance to the hands. (B and D) show what the camera captures in A and C. (E and F) In instructor shots, Stargazer follows the head position of the instructor. }
  \Description{This figure shows body movement cues that Stargazer reacts to. (A and C) In action shots, Stargazer follows the hands of the instructor and adjusts its distance to the hands. (B and D) show what the camera captures in A and C. (E and F) In instructor shots, Stargazer follows the head position of the instructor.}
  \label{fig:cue-hands}
\end{figure}

\subsubsection{Body Movement Cues: Hands}
The vast majority of how-to video shots are shots of the volume in front of the instructor---where their hands are demonstrating or manipulating an object; this is Stargazer's default state.
For such action shots, Stargazer locates both hands of the instructor and tries to: (1) position both near the center of the frame, and (2) identify a proper distance to capture the hands that accommodate for recent hand movement (Figure~\ref{fig:cue-hands} A-D).
Stargazer tracks the midpoint of the two hands and models a spherical volume that contains the hand movement for the past 5 seconds.
This accounts for situations where the distance between the hands changes rapidly---e.g. when the instructor needs to, for instance, select and use tools that are placed adjacent to the object being manipulated.
The camera is positioned such that it looks at the midpoint of the two hands and at a distance where the sphere covers one-third of the width of the camera frame. 

%frame position them around the center of the frame while maintaining a proper framing
%More concretely, the camera tracks the midpoint of the two hands.  
%Stargazer also aims to balance the visibility and legibility of hand motions by finding proper framing, i.e. zooming in when the hands operate in a smaller volume and zooming out when the hand movements are more spread out. 
%As the distance between the two hands can change rapidly when engaged in physical tasks, trying to frame them optimally at everyone moment will lead to abrupt camera motion and unpleasant experience.     
%To follow hands smoothly, we estimate the radius of the smallest sphere that can include both hands based on recent hand positions.    
%\lightGray{notes: talk more about why proper framing is good.}
%Then we position the camera such that it look at the midpoint of the two hands and at a distance where the sphere that encapsulates both hands cover one thirds of the width of the camera frame. 

\subsubsection{Body Movement Cues: Head}
In instructor shots, the face of the instructor is centered horizontally in the frame, and the camera is oriented perpendicular to the instructor's torso (Figure~\ref{fig:cue-hands} E and F).
Following the rule of thirds from cinematography principles~\cite{brown2016cinematography}, Stargazer positions the camera such that the instructor's eyes are at one-third from the top of the camera frame and works to maintain this framing and orientation.

\subsubsection{Body Movement Cues: Torso Orientation}
\begin{figure}[!b]
  \centering
  \includegraphics[width=\columnwidth]{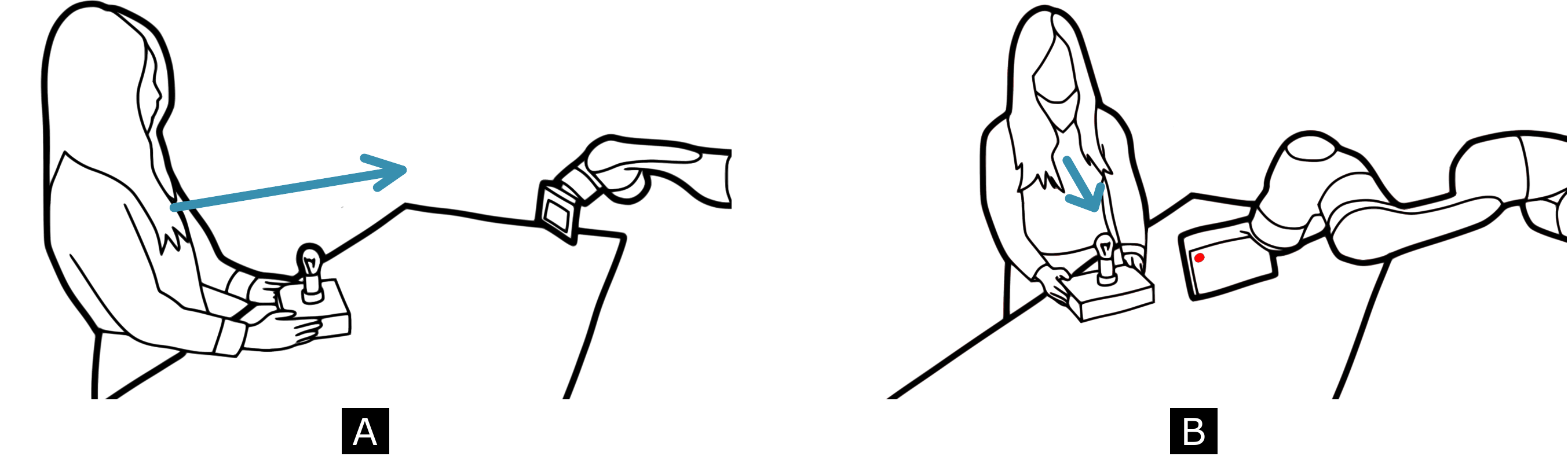}
  \caption{In instructor shots and action shots, Stargazer adjusts its camera orientation and looks perpendicular toward the line connecting the instructor's shoulders. }
  \Description{This figure shows an illustration of how instructor shots and action shots, and stargazers adjust their camera orientation and look perpendicular to the line connecting the instructor's shoulders.}
  \label{fig:cue-torso}
\end{figure}

In instructor and action shots, Stargazer points its camera perpendicularly to the line connecting the instructor's two shoulders to increase subject visibility.
The camera robot reorients itself if an instructor's torso orientation changes (Figure~\ref{fig:cue-torso}).
%for an extended period of time.

\subsubsection{Gestural Cues: Single-Hand Pointing}
\begin{figure}[t]
  \centering
  \includegraphics[width=\columnwidth]{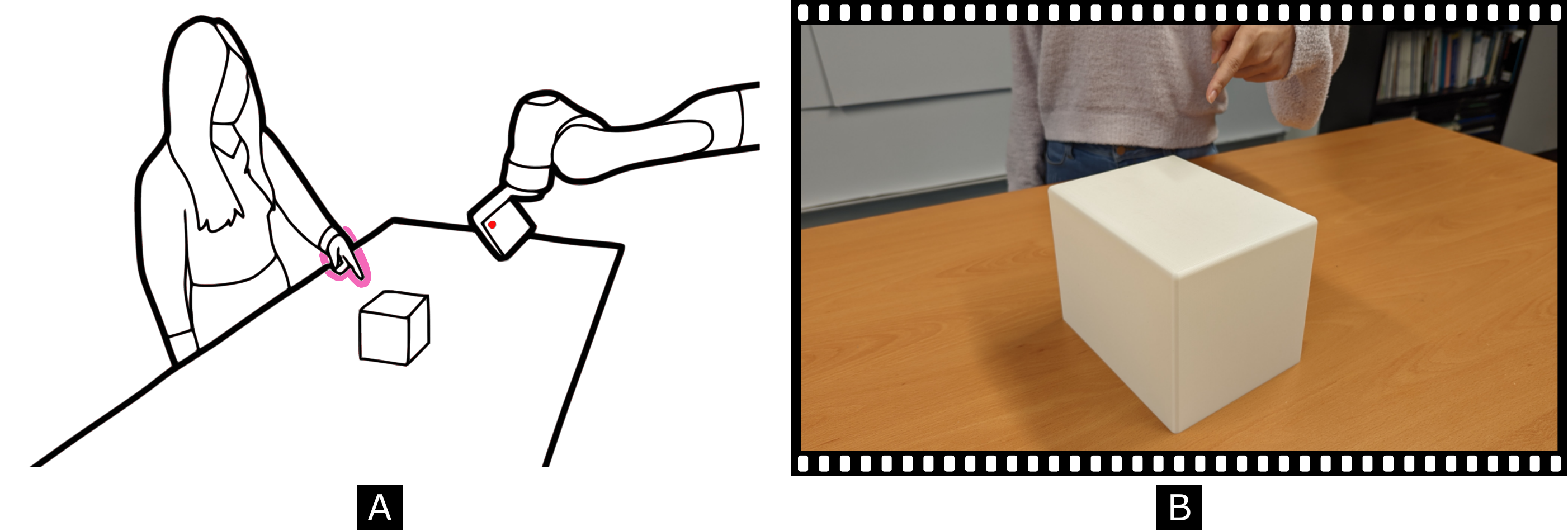}
  \caption{Stargazer transitions to an object shot by following the instructor's pointing hand.}
  \Description{This figure shows an illustration and a photo about how Stargazer transitions to an object shot by following the instructor's pointing hand.}
  \label{fig:cue-point}
\end{figure}

Instructors can direct the attention of the audience by pointing with their index finger at an object---a common communicative action in both everyday life~\cite{gibson1994tools} and instructional settings~\cite{macken2014pointing}.
Stargazer interprets this action as a transition to an object shot by tracking a location that is along the instructor's pointing direction, close to the fingertip (Figure~\ref{fig:cue-point}).
%The instructor points with their index finger to an object to trigger transition to an object shot that focus on this object. 
%The gestures include raising one's hand over the shoulder, and pointing with the index finger.  
%We choose pointing because it is commonly used in everyday communication~\cite{gibson1994tools} as well as instructional settings to direct attention~\cite{macken2014pointing}. 
%They may also use pointing gestures to direct students' attention to an object or its specific component. 
%Such action is well-understood but go unnoticed, as linguist Edward Sapir put, ‘written nowhere, known by none, and understood by all’~\cite{wylie1977guide}.
%Given our focus on tabletop-scale tasks, we assume that the instructor performs proximal rather than distal pointing~\cite{lock1990some}.
%Stargazer's camera tracks the location that is a short distance (12cm in our implementation) away from the fingertip along the pointing direction, as an approximation of the object pointed to. 
Stargazer will continue to follow the instructor's hand while they point (this allows for them to point along a trajectory, for example).
Once the instructor stops pointing, Stargazer returns to capturing action shots, i.e following the hands.

\begin{figure}[t]
  \centering
  \includegraphics[width=\columnwidth]{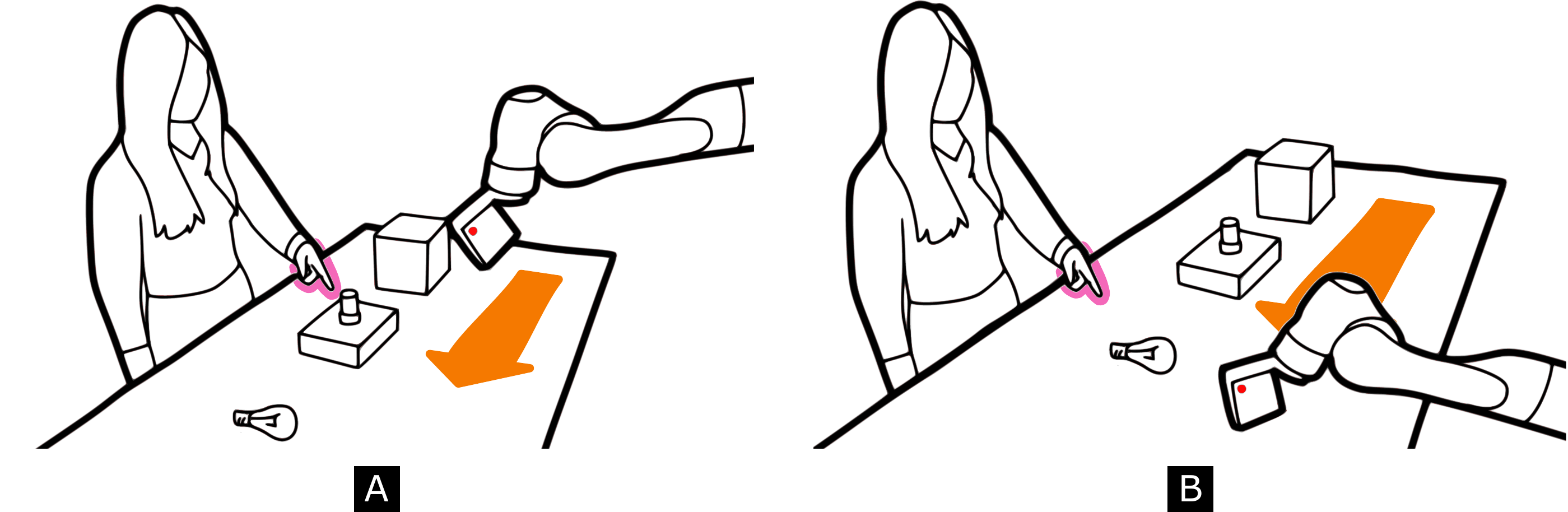}
  \caption{The instructor can move their finger in a horizontal line to trigger a camera ``truck'' effect (i.e., camera moving horizontally along an axis perpendicular to the lens' direction)}
  \Description{This figure shows two illustrations about how the instructor can move their finger in a horizontal line to trigger a camera ``truck'' effect (i.e., camera moving horizontally along an axis perpendicular to the lens' direction)}
  \label{fig:cue-point-truck}
\end{figure}

The instructor can also move their index finger rapidly in a horizontal line while performing the pointing gesture. 
Stargazer interprets this as a signal to perform a linear tracking effect (a ``truck'' movement).
Its camera follows the finger's position while being perpendicular to the finger's trace (Figure~\ref{fig:cue-point-truck}).%, simulating a truck effect.

\begin{figure}[b]
  \centering
  \includegraphics[width=\columnwidth]{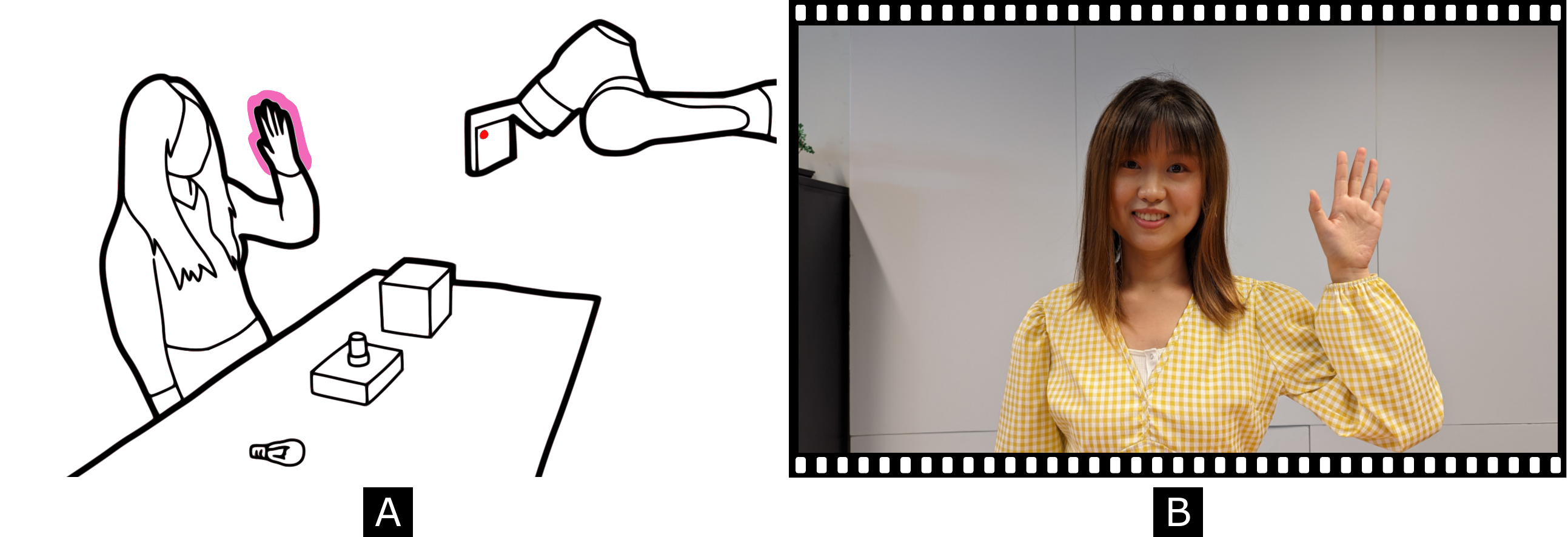}
  \caption{The instructor raises their hand to signal Stargazer to transition to an instructor shot.}
  \Description{This figure shows an illustration and a photo about how the instructor raises their hand to signal Stargazer to transition to an instructor shot.}
  \label{fig:cue-raise-hand}
\end{figure}

\subsubsection{Gestural Cues: Raising Hand}
Raising a hand and waving is a familiar gesture for greeting and mediating attention~\cite{licoppe2017skype}, for example, to open up a conversation.
Stargazer interprets this hand wave signal as a signal to transition to an instructor shot (Figure~\ref{fig:cue-raise-hand}).
%The instructor uses another explicit, however commonly seen gesture in daily social encounters to trigger transition to instructor shots. 
Stargazer specifically recognizes a hand above the shoulder for a short period (1s).
%We align the everyday meaning of the gestures with the purposes of the shots triggered, to hide their other identities as signals to the robot. 
%This alignment could also reduce the instructor's effort to learn and remember to use them.
%Stargazer by default starts with a instructor shot.
%A pointing gesture towards the workbench triggers the transition to an action shot.
%During an action shot, the instructor simply point to an location of interest with her index figure to start an object shot, as she would normally do to highlight an object. 
%Stargazer remains in the object shot mode as the instructor keeps pointing.
%To return back to a instructor shot, the instructor lifts her hand above her shoulder for a short period (1s) as if calling for the attention of the audience.

\begin{figure}[!t]
  \centering
  \includegraphics[width=\columnwidth]{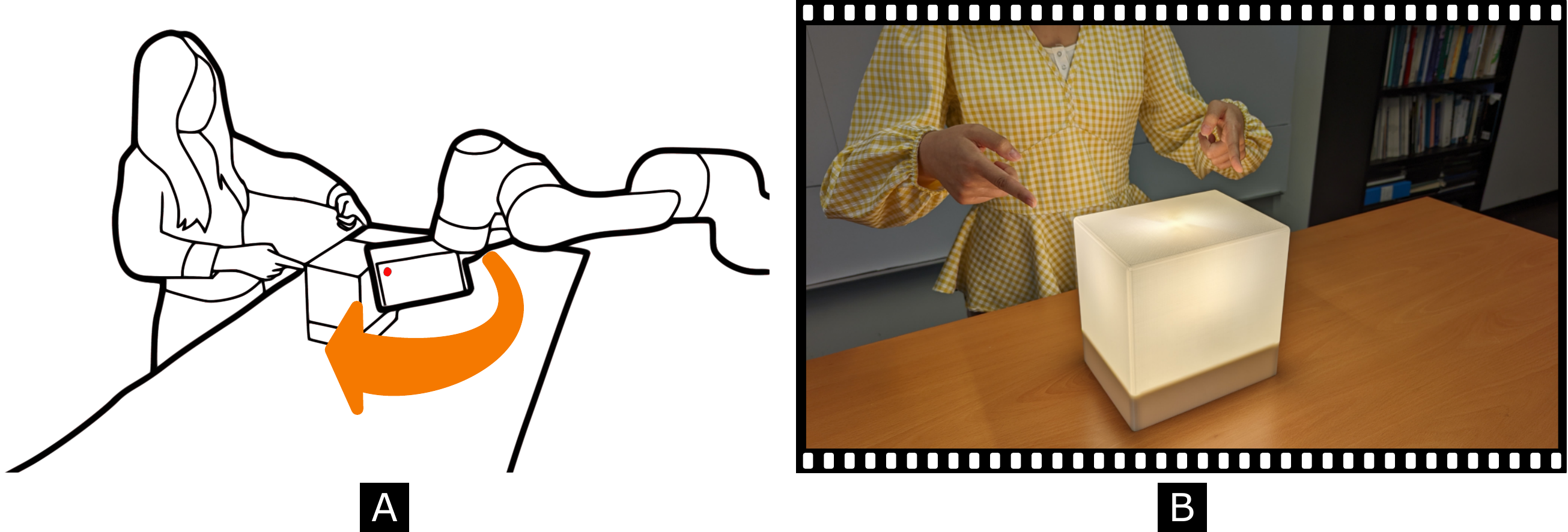}
  \caption{The instructor performs a two-hand pointing gesture to make the camera orbit around the midpoint of the two fingertips.}
  \Description{This figure shows an illustration and a photo about how the instructor performs a two-hand pointing gesture to start an orbit camera movement.}
  \label{fig:orbit}
\end{figure}

\subsubsection{Gestural Cues: Two-Hand Pointing}
The instructor can initiate a camera orbit movement with one finger pointing from each hand as if they are presenting an object to the audience (Figure~\ref{fig:orbit}).  
Stargazer interprets this gesture as a signal to orbiting around the midpoint of the two fingertips, giving an extended shot while the camera moves in an arc.

\subsubsection{Speech Cues: Verbally Suggesting Tight Framing and High-Angle}
During in-person instructional settings, an instructor might suggest to a learner to lean in (e.g., ``If you take a closer look, ...'') or to watch from a different perspective (e.g. ``You can see from the top ...'').
Such speech acts are necessary for motor skill learning~\cite{darden2000revisit}, since they direct the audience to attend to something specific.
Stargazer interprets certain pre-defined speech signals from the instructor as suggestions about camera framing and angle when the dialogue seems to make these recommendations to the audience.
We implemented two specific speech-triggered behaviors.
%Stargazer allows the instructor to control both camera framing and angle through phrases that are embedded in the instruction and seemingly suggest the audience to change their viewpoints.

\begin{figure}[!t]
    \centering
    \includegraphics[width=\columnwidth]{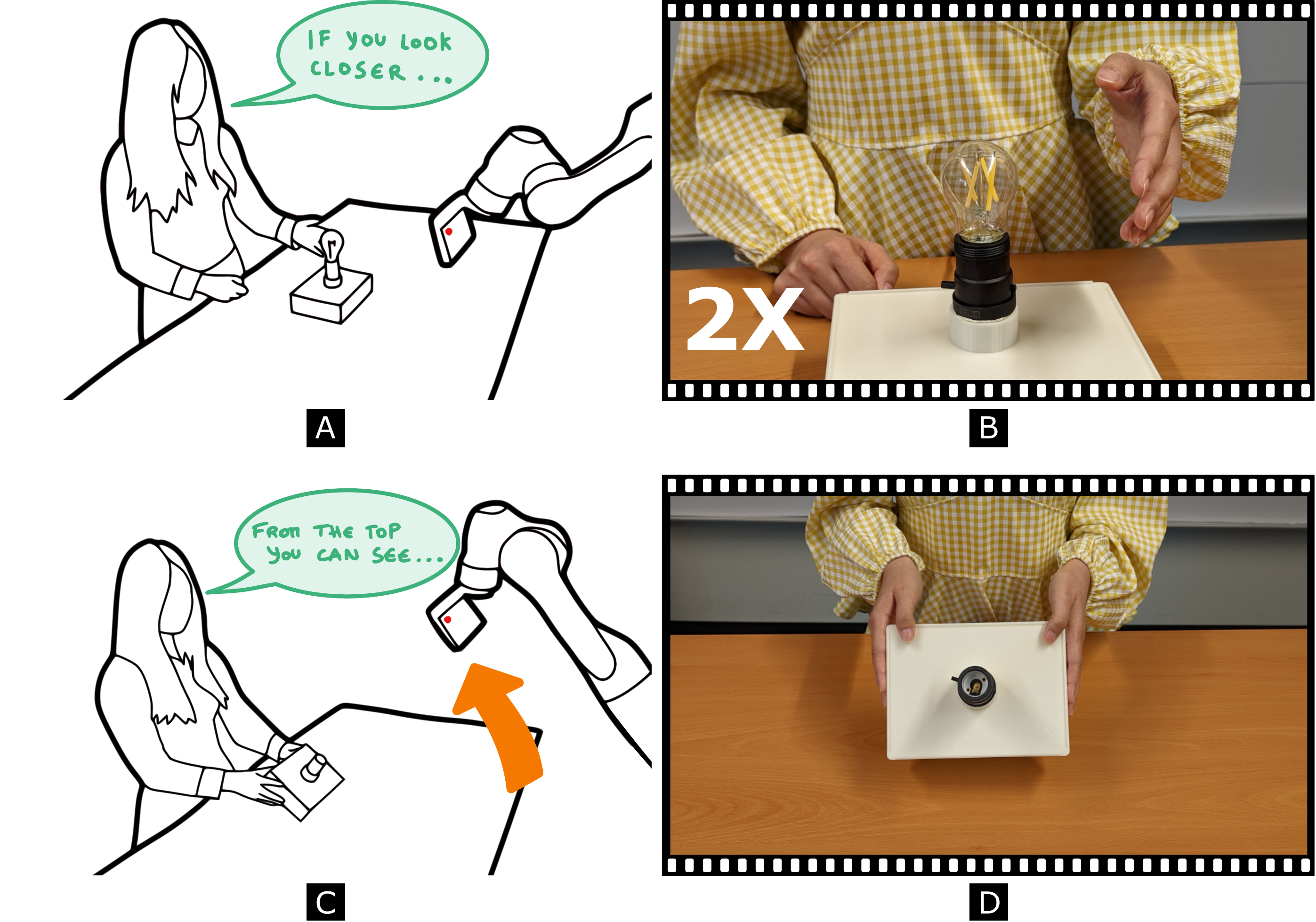}
    \caption{The instructor can (A and B) verbally suggest taking a closer look to make the camera zoom in by 2X, and (C and D) verbally suggest looking from a higher position to take a high-angle shot.}
    \Description{This figure shows two illustrations and two photos about how the instructor can (A and B) verbally suggest taking a closer look to make the camera zoom in by 2X, and (C and D) verbally suggest looking from a higher position to take a high-angle shot.}
    \label{fig:speech}
\end{figure}

Stargazer leverages a state-of-art large language model to infer the intent of the instructor's speech without enforcing strict templates.
%, which allows the instructor to deliver content with domain-specific language.
The instructor can control zoom-in by saying sentences that suggest taking a closer look. For example: ``If you look closer, you can see this socket takes a hexagon shape'', or ``Pay more attention to how I take the lid off.'' (Figure~\ref{fig:speech} A and B).
Similarly, the instructor can control the camera angle (high-angle shot vs. standard-angle shot) with sentences that suggest viewing from a higher position. For example:  ``It is better to look from the top to see how I take the headband off'', or ``I want you to take a top-down perspective now so that you see the full model.'' (Figure~\ref{fig:speech} C and D).
To reset the camera angle to be standard or framing to be normal, the instructor hides one of their hands behind their body.  

%We choose speech input first because instructors often need to explicitly guide learners' attend to specific aspects of actions in motor skill learning~\cite{darden2000revisit}, therefore the use of language to suggest viewpoint change could fit naturally into instruction content.
%\lightGray{Need better references and arguments here.}

Stargazer's use of speech in this way frees the instructor's hands, which are often occupied in how-to videos.
%We use speech for framing and angle control since these may be desirable during skill demonstration, when the hands are occupied.
This is markedly different from our use of gestures for shot transitions involving a shift of subject.
Those transitions typically occur before/after a complete step when an instructor can choose to empty their hands.

\subsection{Feedback to Instructors} \label{sec:feedback}
Prior research suggests the benefits of providing feedback about an autonomous agent's states to people who interact with it~\cite{ju2015design,sheridan1986human}. 
Our early pilot tests also show that instructors need to know what the camera ``sees'' and be able to understand Stargazer's state and near-future actions.
Stargazer provides this feedback with a display attached to the camera mount, which faces the instructor (Figure~\ref{fig:feedback}).
%To inform instructors about what the camera currently captures and the status of the robot, we attach an additional display, which faces the instructor, to the robot camera mount for providing feedback.
The display shows a real-time camera preview, and additionally, an icon for the current shot type, and two indicators showing whether a pointing gesture is detected on the left and right hands.   

\section{Implementation}
\label{sec:implementation}

\begin{figure*}[t!]
  \centering
  \includegraphics[width=0.8\textwidth]{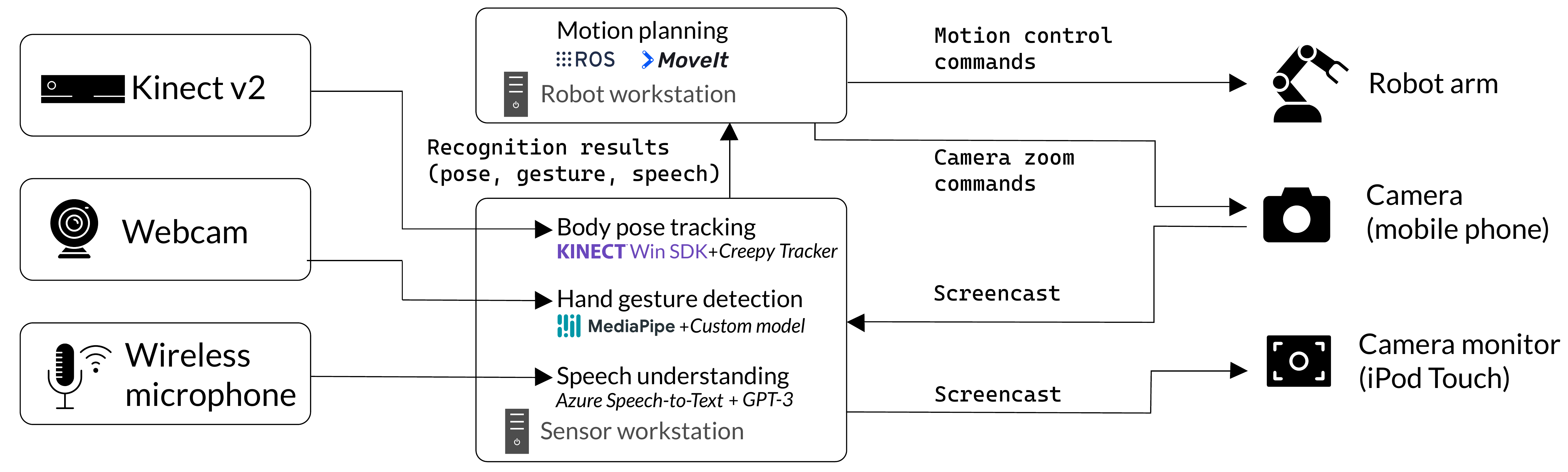}
  \caption{\blue{An overview of the components of the Stargazer prototype.}}
  \Description{This figure shows an overview of the components of the Stargazer prototype, including sensors, workstations and the main software programs running on them, and the robot.}
  \label{fig:prototype-overview}
\end{figure*}

The Stargazer prototype consists of three main parts: the robot arm, the camera, and the sensors (Figure~\ref{fig:prototype-overview}).     
The robot arm is a seven-degree-of-freedom Franka Emika Panda, controlled by a collection of Robot Operating System (ROS) nodes running on a Linux workstation (the robot workstation). 
The sensors include a Kinect v2 depth camera, an RGB webcam, and a wireless microphone for sensing the instructor's skeleton, pointing gestures, and speech. All sensors connect to a Windows workstation (the sensor workstation), which processes sensor data and sends them to the robot workstation \blue{through a wired local network in UDP}.
The camera and the camera monitor are both attached to the robot through a custom mount.
\blue{The software and hardware design of our prototype is open-sourced online\footnote{https://github.com/jchrisli/stargazer-chi23}.}

\subsection{Camera Hardware and Software}
\begin{figure}[t]
    \centering
    \includegraphics[width=\columnwidth]{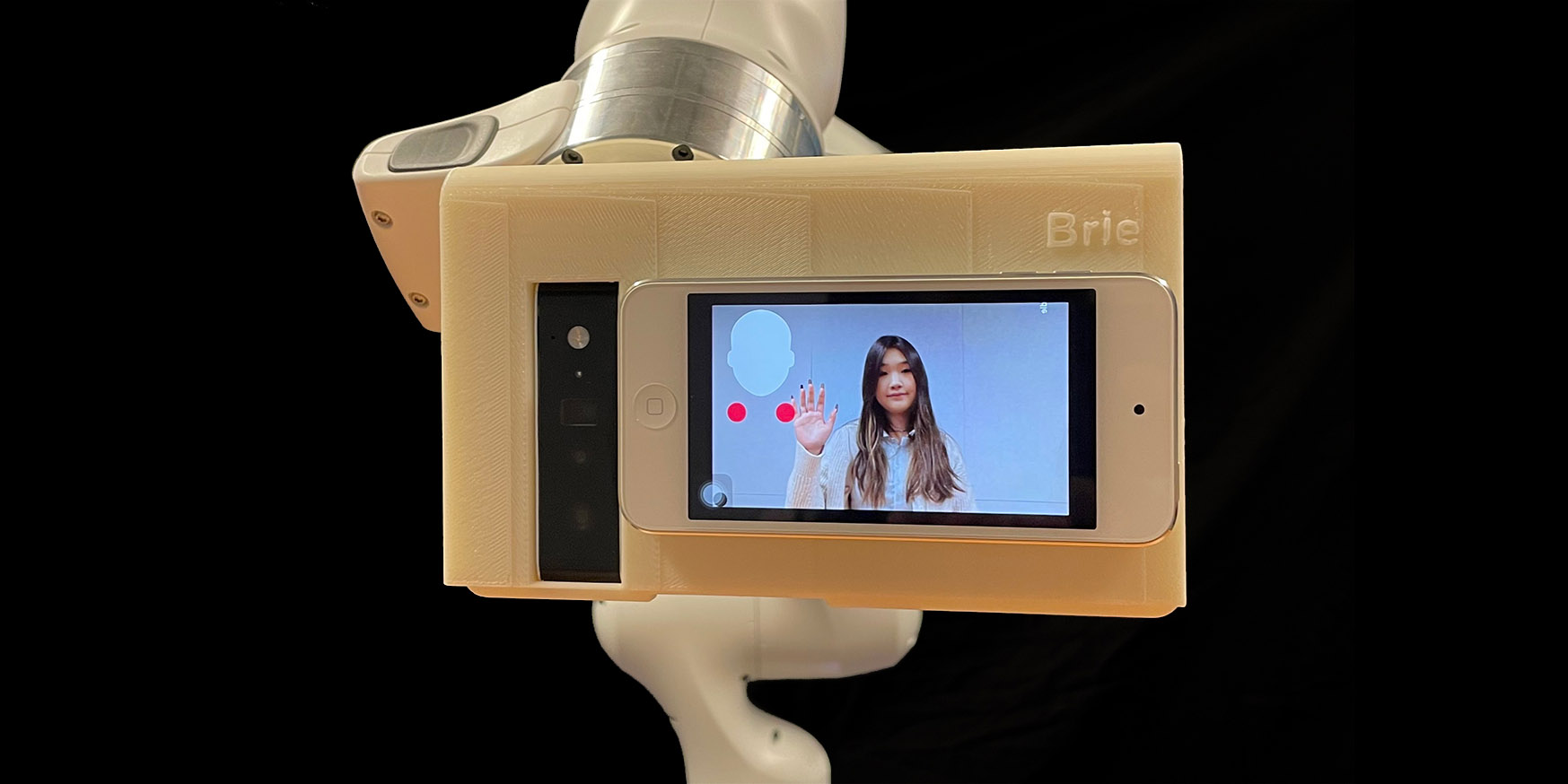}
    \caption{The camera monitor provides feedback to the instructor, showing the camera preview, the current shot type, and whether a pointing gesture is detected. In this photo, the monitor shows a head icon indicating the current shot type is an instructor shot, and two red dots indicating neither hand is pointing.}
    \Description{This figure shows a photo of the camera monitor providing feedback to the instructor, showing the camera preview, the current shot type, and whether a pointing gesture is detected. In this photo, the monitor shows a head icon indicating the current shot type is an instructor shot, and two red dots indicating neither hand is pointing.}
    \label{fig:feedback}
\end{figure}

We use the back-facing camera of an Android mobile phone (Google Pixel 6 Pro) as the camera of Stargazer.
\blue{The phone runs a custom software application to record video and communicate with the robot workstation through a wireless UDP connection.
The application shows the preview of the phone's camera feed and several widgets showing the current shot type and gesture detection results, as described in Sec.~\ref{sec:feedback}.
The robot workstation sends messages to this application to control the camera's zoom level and update the status widgets.
The camera monitor mirrors the screen of the phone to provide feedback to the instructor (more details in Sec~\ref{sec:monitor}).}
The phone is inserted into a 3D-printed mount (Figure~\ref{fig:feedback}), which is attached to the end of the robot arm as a custom end-effector.

\subsubsection{Camera Monitor} \label{sec:monitor}
An iPod Touch is mounted to the instructor-facing side of the camera mount as the camera monitor.
It displays the preview of the mobile phone's camera feed and robot status, mirroring the phone's screen \blue{through a local wireless network}.
We use the sensor workstation as a relay point for screen mirroring.
The sensor workstation first obtains a screencast of the phone through scrcpy\footnote{https://github.com/Genymobile/scrcpy} and streams the screencast to the iPod Touch via Moonlight\footnote{https://moonlight-stream.org/}.

\subsection{Sensing Instructor's Actions}
The Stargazer prototype employs a combination of sensors to detect the body pose, gesture, and speech of the instructor. 
Future work can explore leveraging the sensing capabilities of the robot's camera itself.

\subsubsection{Body Pose}
We point a Kinect v2 depth camera at the instructor and infer the instructor's body pose using the Kinect SDK for Windows.
Stargazer uses the head joint as a proxy for the face, the fingertip joints for the hands, and the vectors from the wrist joints to the corresponding fingertip joints for pointing directions.
In object shots, we track a location that is 12cm from the fingertip along the pointing direction. 
To estimate the spherical volume that contains both hands, we run a rolling average of the distance between the hands for the most recent 5 seconds.    
%We calculate the torso orientation by computing the cross product of the vector from one shoulder joint to another and the gravity vector.
For filtering and streaming the data, we used the Creepy Tracker toolkit~\cite{sousa2017creepy}.

\subsubsection{Hand Gesture}
\blue{
A 720p RGB webcam is aimed at the center of the workbench to detect the instructor's pointing gestures.
} 
We locate the 2D keypoint positions of the instructor's hands using a deep-learning-based detector~\cite{lugaresi2019mediapipe} and classify whether \blue{the hand is performing a pointing gesture (index finger extending forward while middle, ring, and pinky curled up)} through a custom model.
The 21 key point positions from each hand have been normalized before classification.
\blue{The gesture detection model is a four-layer fully-connected neural network implemented with Keras, trained with data collected from the research team.}
\blue{To reduce ambiguity in classification, we only classify the current gesture as pointing if the model has at least 85\% confidence.
We further apply low-pass filtering to remove sporadic false positives.}

\subsubsection{Speech Understanding}
We record the instructor's speech with a wireless microphone clipped to the instructor and send it to a speech recognition API (Microsoft Azure Speech-to-Text\footnote{https://azure.microsoft.com/en-us/services/cognitive-services/speech-to-text/}).
The transcribed text, along with a custom prompt\footnote{We attach the prompt we used in the supplemental materials.}, is then sent to a large language model (GPT-3~\cite{brown2020language}), which labels the intent (i.e., tighter framing/high angle/normal) of the instructor.
% To reduce latency, we send interim speech recognition results for labeling.

\subsection{Robot Control}

The Stargazer prototype uses ROS to communicate with the robot arm and control its motion.
The robot workstation runs a ROS node that computes the next target camera pose based on sensor inputs. 
We describe the algorithm for planning the next camera poses in Section~\ref{sec:motion-planning}.
Given a target camera pose, we get the desired Cartesian linear and angular velocities with a PID controller.  
A real-time servoing library, MoveIt Servo, converts the Cartesian velocities to joint velocities.
Finally, we set the joint velocities through the Franka ROS interface\footnote{https://frankaemika.github.io/docs/index.html}.  
The next camera pose is calculated at a frequency of 5 Hz. 
The PID control loop runs at 100 Hz.

We monitor the robot's status to determine if it is close to kinematic singularities or joint limits which would render it unable to move properly.
If so, the control program pauses real-time velocity control and recovers the robot from the error state by returning to a safe neutral position.

\subsection{Robot Motion Planning} \label{sec:motion-planning}

\begin{figure}[b]
  \centering
  \includegraphics[width=\columnwidth]{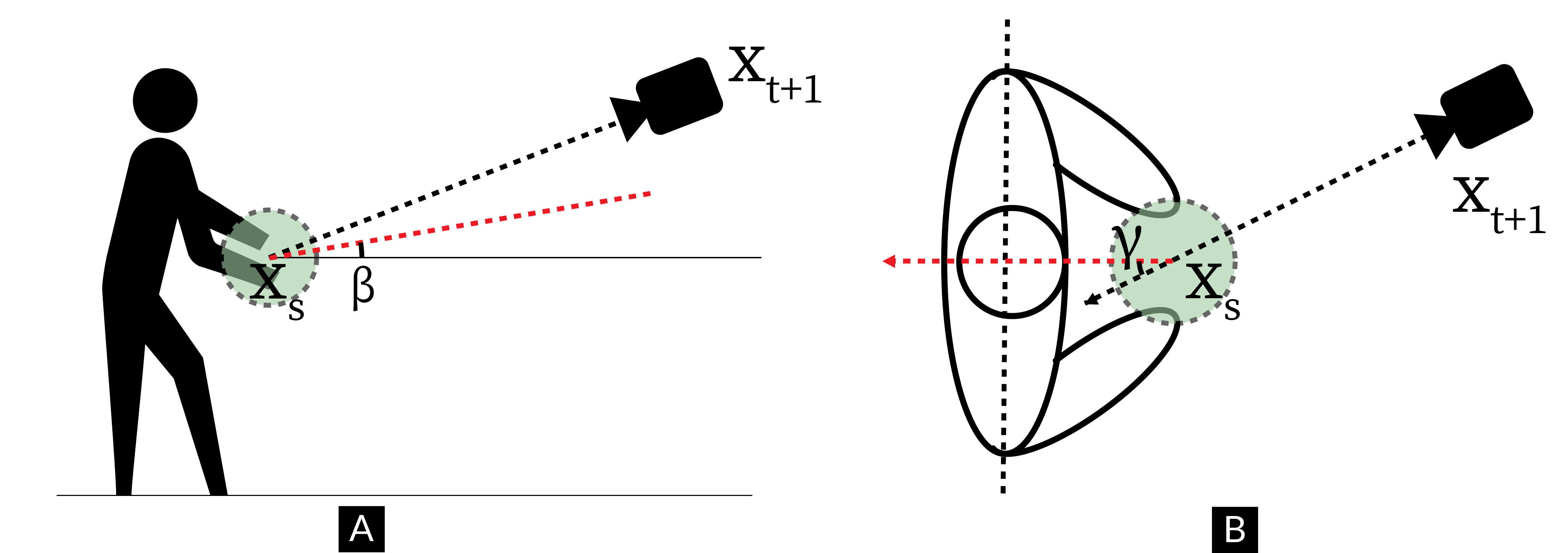}
  \caption{(A) The cost for the difference between the camera pitch angle of the next camera position $\mathbf{x_{t+1}}$ and the desired pitch angle (in red) $\beta$. (B) The cost for the desired camera orientation (in red) and the orientation of the next camera position $\mathbf{x_{t+1}}$. $\gamma$ indicates the difference. }
  \Description{This figure shows two illustrations for the cost for the difference between the camera pitch angle of the next camera position and the desired pitch angle and the cost for the desired camera orientation and the orientation of the next camera position}
  \label{fig:path-planning}
\end{figure}

While the camera pose has six degrees of freedom, Stargazer's camera always aims to look at a known location (e.g. the instructor's head) and keep its image plane's horizontal axis parallel with the ground. 
This reduces our motion planning problem from $\mathbb{R}^6$ to $\mathbb{R}^3$.
We aim to plan the next target camera position $\mathbf{x_{t+1}} \in \mathbb{R}^3$ for capturing the subject (e.g., face, both hands) at position $\mathbf{x_{s}} \in \mathbb{R}^3$ under a number of constraints, such as motion smoothness and distance to the subject, at time step $t$.       
Except for planning the orbit paths, we formulate the next position selection as an optimization problem, following prior work in autonomous drone cinematography~\cite{nageli2017realtime}.
We calculate the orbit path directly as a series of waypoints, as described in Section~\ref{sec:orbit-paths}.  
%To simplify the bounds on the variables, we parameterize the camera position $x$ as $(\theta, \phi, R)$ in a polar coordinate system with its origin at the second joint of the robot.
\subsubsection{Motion Planning Cost Function} \label{sec:cost-function}
The cost function $J$ consists of the weighted sum of the following cost values.  

\noindent\textbf{{Motion Smoothness:}} 
To achieve smooth camera motion, we penalize large displacement in the camera position with the cost $c_{ms} = (\mathbf{x_{t+1}} - \mathbf{x_{t}})^{2}$. 

\noindent\textbf{Desired Distance:}
We first compute the desired distance $d$ to film the subject based on the current type of shot.
For action shots, the subject should cover one-third of the camera frame's width.
Therefore, for a subject with radius $r$ and a camera with field-of-view $\alpha$, the desired distance $d=2r\tan{(\alpha/2)}/3$.
For instructor shots and object shots, we empirically set $d$ to be 0.20m and 0.12m, respectively.
Given $d$, the desired-distance cost is $c_{dd}=(\|\mathbf{x_{t+1}}-\mathbf{x_{s}}\| - d)^2$.
We do not enforce distance cost for high-angle shots as it tends to drive the robot into kinematic singularities. 

\noindent\textbf{Pitch:}
For instructor shots and shots with a high-angle setup, we encourage a camera pitch value close to $\pi/2$ and 0, respectively. 
We calculate a pitch cost for the desired pitch value $\beta$ as $c_{p}=((\mathbf{x_s}-\mathbf{x_{t+1}}) / \|\mathbf{x_s}-\mathbf{x_{t+1}}\| \cdot \mathbf{v_g} - \arccos{(\pi/2-\beta)}) ^2$, where $\mathbf{v_g}$ is the gravity direction $(0, 0, -1)$ (see Figure~\ref{fig:path-planning}A). 

\noindent\textbf{Orientation:}
We add a camera orientation term $c_o$ in two conditions. 
For face and action shots, we encourage the camera to look perpendicularly at the line linking the instructor's two shoulders.
During truck movements, we guide the camera to point perpendicularly at the finger's movement direction. 
Since we already have a cost term for controlling camera pitch, we only need to penalize the difference between the camera orientation vector of the next step and the target camera orientation vector in the ground plane (Figure~\ref{fig:path-planning}B).
%Let the target camera orientation and the camera orientation of the next step projected in the ground plane be $\mathbf{v_o}$ and $\mathbf{v_{t+1}}$, respectively, the orientation cost can then be written as $c_o=\| \mathbf{v_{t+1}} - \mathbf{v_o} \|^2$. 
We refer our readers to Appendix.~\ref{sec:appendix-orientation} for the detailed derivation of $c_o$.

In summary, the cost function for selecting the next camera position is
\[J=\omega_{sm}c_{sm}+\omega_{dd}q_{dd}c_{dd}+\omega_{p}q_{p}c_{p}+\omega_{o}q_{o}c_{o}\]
where $\omega_{sm}$, $\omega_{dd}$, $\omega_{p}$, $\omega_{o}$ are weights for the cost terms. 
If the current shot is not a high-angle shot, $q_{dd}=1$  otherwise 0.
If the current camera angle is set to high-angle or the current shot is an instructor shot, we set $q_{p}=1$, otherwise 0.
If the current shot is a hand or instructor shot, or the camera is performing a truck movement, $q_{o}=1$  otherwise 0.
For weights, we empirically set $\omega_{sm}=1.0$, $\omega_{dd}=0.2$, $\omega_{p}=1.0$, $\omega_{o}=0.5$ to achieve a balance between responsiveness and smooth tracking. 
To simplify the bounds on the variables, we re-parameterize $\mathbf{x_{t+1}}$ as $(\theta_{t+1}, \psi_{t+1}, R_{t+1})$ in a polar coordinate system.  
We solve this nonlinear optimization problem with the SLSQP solver in SciPy.
We refer our reader to Appendix.~\ref{sec:appendix-polar} for the details of the polar coordinate system used and the bounds on the variables.

\subsubsection{Orbit Paths} \label{sec:orbit-paths}
Given an orbit center, the Stargazer camera always travels along an arc of $\pi/4$ and looks at the orbit center from a distance of 60cm with a pitch angle of $\pi/6$.

\subsubsection{Implementation Details}
\blue{
The robot workstation runs Ubuntu 20.04 and ROS 1 Neotic with an Intel Xeon W2295 CPU and 503 GB memory. 
The sensor workstation runs Windows 11 with an Intel i7 8750 CPU and 16 GB memory.
We implemented one performance-critical ROS node, which calculates end-effector velocities, in C++, and the other nodes in Python.
The speech detection and labeling program were implemented in C\#.
The programs for other sensors were implemented in Python.
}

\section{Evaluation}
We conducted a preliminary user evaluation where six instructors created video tutorials demonstrating distinct physical skills.
The goal of the evaluation was to gain an initial understanding of Stargazer's ability to support the filming of how-to videos and the instructor's perception of working with Stargazer.  

\subsection{Study Design and Participants}

\begin{table}[!b]
\centering
\resizebox{\columnwidth}{!}{%
{\renewcommand{\arraystretch}{1.2}
\begin{tabular}{@{}lll@{}}
%\toprule
\textbf{ID} & \textbf{Background} & \textbf{Skill Taught} \\ \midrule
P1 (Pro) & Artist in digital fabrication. & \begin{tabular}[t]{@{}l@{}}Photogrammetry and 3D\\ printing.\end{tabular} \\
P2 (EH) & \begin{tabular}[t]{@{}l@{}}Skateboarding influencer, running \\ an Instagram channel teaching \\ skateboarding skills.\end{tabular} & Skateboard maintenance. \\
P3 (EH) & Virtual reality researcher. & \begin{tabular}[t]{@{}l@{}}Head-mounted display \\ set-up.\end{tabular} \\
P4 (EH) & Crafting hobbyist with art training. & Clay model making. \\
P5 (Pro) & \begin{tabular}[t]{@{}l@{}}Filmmaker, teaching stop-motion \\ filmmaking to college students and \\ the general public.\end{tabular} & \begin{tabular}[t]{@{}l@{}}Stop-motion puppet \\ making.\end{tabular} \\
P6 (Pro) & Digital artist and filmmaker. & \begin{tabular}[t]{@{}l@{}}Interactive sculpture \\ making.\end{tabular} \\ %\bottomrule
\end{tabular}%
}
}
  \caption{Evaluation Participants (Pro - Professional; EH - Experienced Hobbyist)}
  \label{tab:participants}
\end{table}

\begin{comment}
\begin{table*}[!b]
\resizebox{\columnwidth}{!}{
\begin{tabular}{@{}lllll@{}}
    \textbf{ID}  & \textbf{Background} & \textbf{Skill Taught}\\
    \midrule
    P1 (Pro) & Artist in digital fabrication. & Photogrammetry and 3D printing. \\
    P2 (EH) & Skateboarding influencer, running an Instagram channel teaching skateboarding skills. & Skateboard maintenance. \\
    P3 (EH) & Virtual reality researcher. & Head-mounted display set-up.\\
    P4 (EH) & Crafting hobbyist with art training. & Clay model making.\\
    P5 (Pro) & Filmmaker, teaching stop-motion filmmaking to college students and the general public. & Stop-motion puppet making.\\
    P6 (Pro) & Digital artist and filmmaker. & Interactive sculpture making.\\
    %\bottomrule
  \end{tabular}
  }
  \caption{Evaluation Participants (Pro - Professional; EH - Experience Hobbyist)}
  \label{tab:participants}
\end{table*}
\end{comment}

We invited six physical skill instructors (four female, two male) to create video tutorials for skills that they have expertise in.
Three participants (P1, P5, and P6) were artists who practiced the craft explained in their tutorials professionally. 
The other three participants were experienced hobbyists with more than a year of experience in the skill they taught.
Two participants (P2 and P5) had experience teaching physical skills through online videos or live streaming.
P5 and P6 were professional filmmakers. 
See Table.~\ref{tab:participants} for more details on the participants.
Participants were compensated for their time.

The study protocol has been approved by the ethics review board of our institutions.

\subsection{Task and Procedure}
\begin{figure}[!t]
  \centering
  \includegraphics[width=\columnwidth]{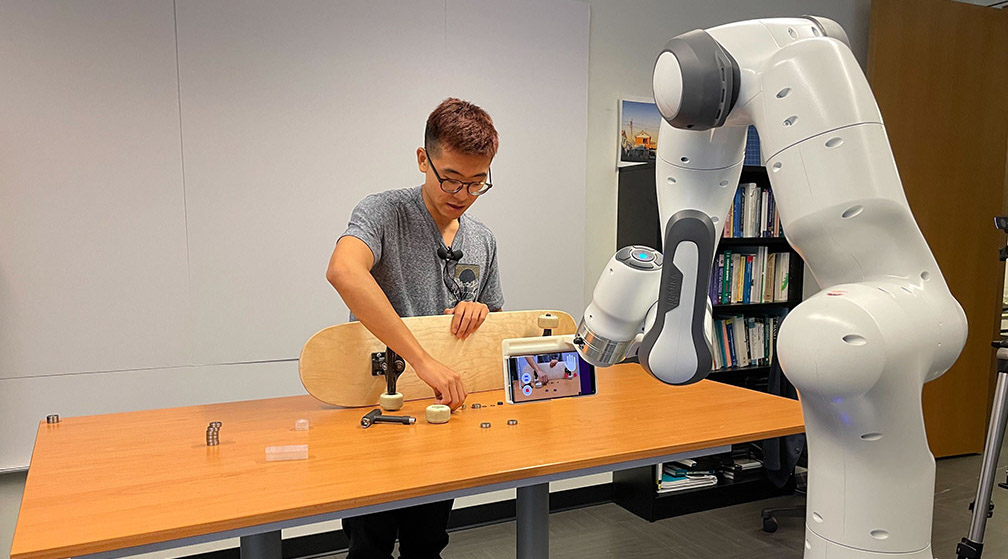}
  \caption{Participant recording a how-to video on skateboard maintenance. }
  \Description{This figure shows a photo of a participant recording a how-to video on skateboard maintenance with Stargazer.}
  \label{fig:participant}
\end{figure}

%We designed the study task in such a way that participants could create 
We asked the participants to freely demonstrate their physical skills without constraints. 
The only limitations were the scale (the task was demonstrated on a workbench) and the length of the video instruction (around 5 minutes per session). 
Participants brought their own materials.
Figure~\ref{fig:participant} shows a participant filming a video with Stargazer.
Each session lasted around 90 minutes and consisted of three phases:

\noindent\textbf{Training.} After filling out a demographics survey, participants went through a training session where they learned about the features of Stargazer, including the types of shots it offers and the interactions it supports for changing camera behaviors.  
They experimented with all the features and recreated a short Lego assembly tutorial video, which required using each feature at least once.

\noindent\textbf{Rehearsal and Review.} After participants successfully recreated the Lego tutorial, they filmed a rehearsal version of their own physical skill tutorial with Stargazer.
The experimenter did not provide any guidance or feedback during filming. 
The participants then reviewed the rehearsal version with the experimenter, using think-aloud to remark on the video and reflect on which parts could be improved.
We included the rehearsal stage for participants to practice delivering their instruction.

\noindent\textbf{Filming and Review.} Participants then created a second take of the tutorial video, incorporating changes in content and delivery they had planned after reviewing the rehearsal version.  
Again, participants used think-aloud to review the outcome and comment on the quality of the video and their experience when filming it. 
% The experimenter probed participants further if their comments were relevant to the behavior of the robot and instructor signals.
Following the review phase, the experimenter conducted a semi-structured interview with participants, focusing on alternative ways to film the videos, how the robot could fit into their current tutorial creation workflow, and comparing Stargazer with existing video capturing solutions for real-time transitions between multiple camera sources, such as Open Broadcast Studio (OBS).
Finally, participants answered a questionnaire consisting of 7-point Likert scale questions about their perception of the robot's behaviors, the interactions, and the video they produced.
We recorded robot motion and user input logs for preliminary performance analysis.
\blue{As a first step in prototyping interactive camera robots for filming how-to videos, our data collection focused on the perspectives of instructors to get their firsthand account of how Stargazer's interaction design might affect instruction delivery and video quality.
Future work could instead study the perspectives of external experts and learners on produced videos.}

\begin{figure*}[h]
  \centering
  \includegraphics[width=0.9\textwidth]{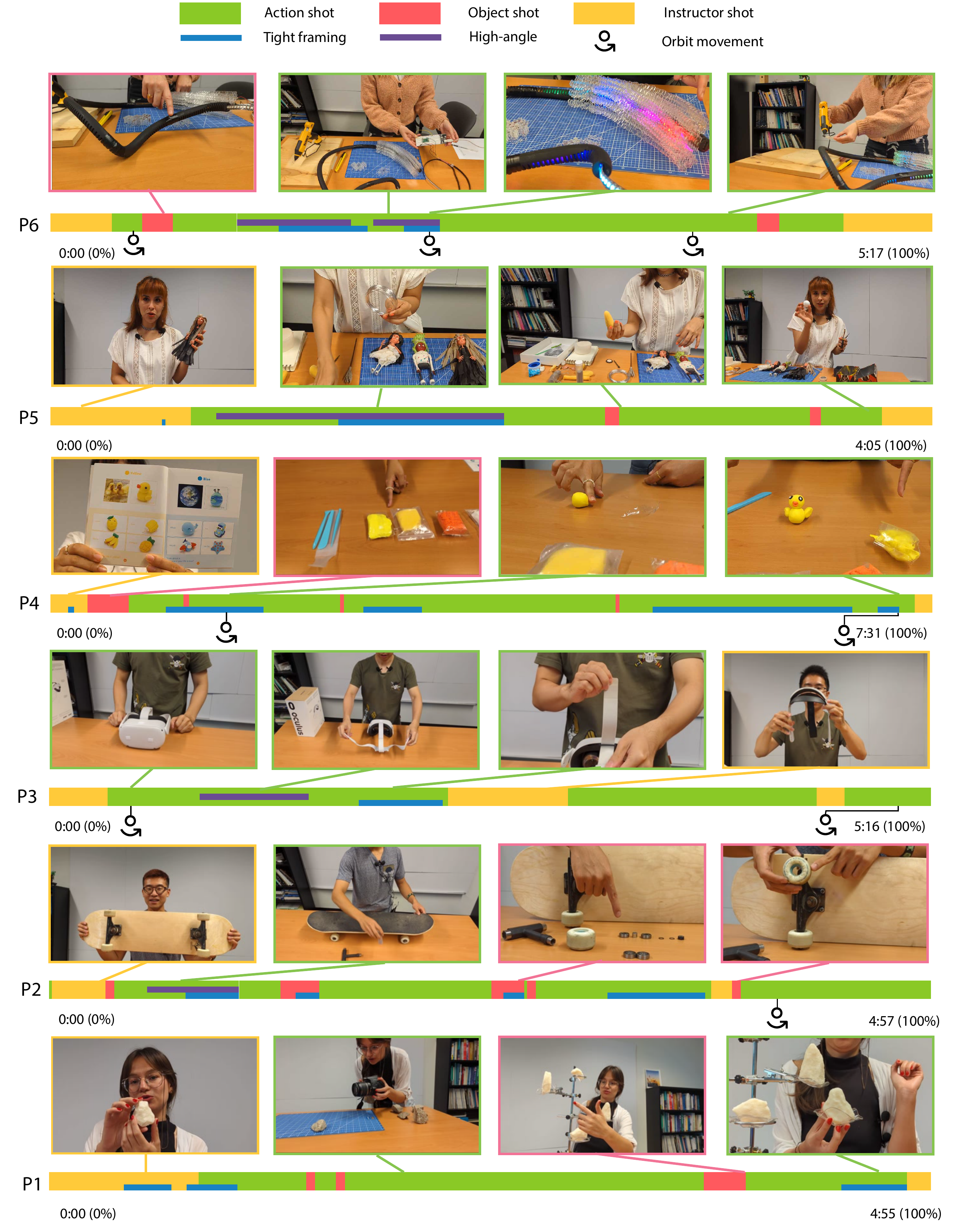}
  \caption{Timelines with the videos' participants filmed with Stargazer. The timelines are color-coded for the shot type, camera angles, camera framing, and camera movements. The representative frames for each video are above the timelines.  }
  \Description{This figure shows a number of timelines for the video's participants filmed with Stargazer. The timelines are color-coded for the shot type, camera angles, camera framing, and camera movements. The representative frames for each video are above the timelines. }
  \label{fig:video-timeline}
\end{figure*}

\subsection{Setup and Apparatus}
We ran the study in a controlled lab environment.
The workbench where the participants demonstrated their skills was 1.6 meters x 0.8 meters, placed 1.2 meters from the base of the robot.
We chose this physical setup and placed constraints in motion planning so that the robot was distant enough to avoid any collision between the robot and the instructor.
One researcher guided participants through the experiment and conducted the interview. 
The setup of the robot is as described in Section~\ref{sec:implementation}.

\subsection{Results}

In this section, we describe the video results created by participants, their perception of interacting with Stargazer, and preliminary technical performance measures.

\subsubsection{Training}
All participants were able to successfully use all the features of Stargazer while filming the Lego assembly tutorial in the training session.
Participants picked up gesture-based interactions very quickly but it took them 2-3 trials to learn to add speech commands into instruction content.

\subsubsection{Video Content}
All participants were able to complete their tutorial videos after the first (rehearsal) or second take. 
Two participants (P4 and P6) were satisfied with the results of their first takes and stopped filming. 
The remaining participants did a second take.  
Our following analysis of the video content is based on the final versions of the video results.     

Figure~\ref{fig:video-timeline} illustrates keyframes and timeline summaries of each of the videos created by participants.
The lengths of the final videos ranged from 245 seconds (4m 5s) to 451 seconds (7m 31s) (M=320s, SD=69s). 
The skills in each video varied significantly in terms of the artifacts and processes required, and participants imprinted their personal style in the delivery of the content.     
Overall, 16.1\% of the total video duration featured instructor shots, 78.4\% was action shots, and 5.5\% was object shots.
Participants incorporated a rich variety of shots by leveraging different combinations of subjects, camera framing, and camera angles (see Figure~\ref{fig:video-timeline}). 
On average, each video had 7.7 times of subject changes (e.g., from an action shot to an object shot, SD=2.6), 5.7 framing changes (e.g., zoom-in, SD=2.9), and 1.7 angle changes (e.g., from standard to high angle, SD=1.5).
Participants were also able to supplement their videos with camera movements, albeit to varying extents. 
They used orbit movements 1.2 times per video on average (SD=1.2).
There was only one instance of a participant using the truck movement.
We attach the participant-produced videos in our supplemental materials.

\subsubsection{Video Quality}
Participants reviewed the tutorial videos they produced and remarked on the quality of the results.   
All participants reported they were able to successfully deliver the instruction content that they had planned in the produced videos (M=6.5, IQR=0.75).
Additionally, they found the videos informative (M=6.3, IQR=1.0), with good visual quality (M=6.5, IQR=0.75) and cinematographic quality (M=6.0, IQR=1.5).
One participant (P6), who is a professional filmmaker, highlighted the visual characteristics of ``one-take'' videos, which Stargazer films, ``\textit{there is interesting theory attached to it (one-take films) about continuity and slowness that allows the viewer to be fully present. }''.
\blue{Indeed, professionals have long valued cinematic continuity in film editing and research in cognitive science posits that continuity helps with guiding viewers' visual attention~\cite{smith2012attentional}. 
An interesting question for the fields of HCI and educational psychology is how cinematic continuity in learning materials may affect physical skill learning.   
}

\subsubsection{Participants' Overall Perception}
Participants found Stargazer useful for creating tutorial video content (M=6.3, IQR=0.75).
The professional artist participants (P1, P5, P6) stated that they often need to document their work processes through video, ``\textit{we ask our colleagues to help (filming) but they are not always available}'' (P1).
When comparing Stargazer to their current documentation workflow, they appreciated that it could alleviate the burden of setting up cameras and post-editing, ``\textit{do my thing and it (filming) just happens}''~(P5).   

\subsubsection{Robot Behaviors and Instructor Control}
%how well our prototype supported creating the content they want to make
%how well the interaction worked for them. Do they fit in their workflow? Are they expressive enough? What does not work?
We conducted an in-depth analysis into how the behaviors of Stargazer and the associated interactions worked for participants, based on a synthesis of interview data, questionnaire responses, and their think-aloud comments while reviewing the captured videos. 

\noindent\textbf{Shot types.} All participants found the three types of shots--action, instructor, and object--sufficient for their purposes when creating the tutorials (M=6.0, IQR=1.5).
As expected, participants mostly used instructor shots to introduce themselves, the task, and certain objects.
They used action shots for illustrating the steps and used object shots to highlight objects and details on objects.  
However, we also observed that participants were able to flexibly adapt the robot's behaviors for their in-situ filming needs and blurred the boundaries between the anticipated purposes of the shots.
For example, in her tutorial on stop-motion puppets, P5 often described a component while holding it close to her chest. 
This hand position leads to the camera capturing both her face and hands while the robot is still in the mode of filming action shots.
She explained that she guided the robot's camera into this position as she wanted to talk to the audience face-to-face more often.

\noindent\textbf{Camera framing and angles.}
Most participants (5/6) reported that they were able to film the tutorial content with the framing and angle they desired (M=5.3, IQR=2.5 for framing, M=5.1, IQR=1.0 for angles). 
All participants see clear utility in the ability to adjust framing (M=6.6, IQR=0.75) and angles (M=6.5, IQR=1.0) using speech.     

Participants mostly used normal (medium for instructor shots, close-up for action and object shots) framing (78.9\% of the total video duration), but switched to tighter framing to highlight details. 
Control of framing especially benefited tasks that involved more actions and objects at a fine scale.
P4 applied tighter framing for 43\% of her full video to ensure that individual parts of the clay model, including smaller components like facial features, were clearly visible to the audience (e.g. the last keyframe on P4's timeline in Figure~\ref{fig:video-timeline}). 
%Similarly, P2 used tighter framing for showing the process of install the tiny pieces on a skateboard  
Participants used high-angle shots to show an overview of the workbench (e.g., P2's second keyframe on Figure~\ref{fig:video-timeline}) or to take a different, visually interesting perspective of the same object (e.g., P5, P6, the second keyframes on both of their timelines in Figure~\ref{fig:video-timeline}).   
After showing how individual components constitute a sculpture, P6 made the robot look downward to show the sculpture's complete structure.

\noindent\textbf{Blending Camera Control into Instruction.} 
Overall, participants found that they could blend camera controls fluidly into instructional activities.
One participant (P5) drew on her experience teaching live remote art classes via a video recording tool (Open Broadcaster Software, or OBS), which supports real-time switch between video sources , and stated that Stargazer's controls did not force her to ``\textit{stop my work and reach for that box (a keypad for changing camera sources in OBS) every time}''.

Four participants explicitly commented on the ease of using the pointing gesture to initiate object shots, finding that it naturally fit into their actions without being disruptive to instruction (M=6.7, IQR=0.75), as P6 described, ``\textit{do not need to think about it}''.  
P4 liked that she could quickly shift the camera between a tool and the work at hand through pointing.  
However, P1 found that the transient nature of the pointing gesture somewhat limited her freedom to construct an object shot.
To make the camera lock onto a particular object while she was fetching other tools, she had to keep pointing at the object with one hand and grab the tools with her other hand.
She suggested that the robot should ``\textit{keep focusing on an object unless I tell it to stop}''.

Participants reported that raising their hand to trigger instructor shots did not disrupt their flow but was less natural than pointing (M=5.2, IQR=2.25).
P3 explained ``\textit{it (raising a hand) feels very natural when you have something to show in you hand but not so much if you just want to quickly talk to your audience}''.
P6 suggested the potential of gaze, ``\textit{I'd like it to turn to my face if I look at it, and go back to my hands if I look away}''. 

Five out of the six participants found controlling camera framing and angles with speech did not disrupt their instruction delivery (M=5.5, IQR=1.75).
P2 considered it his favorite feature, as ``\textit{I can call it to zoom in anytime even if both of my hands are occupied...Being able to show the details makes me want to explain more because they (the audience) can see it.}''
Participants liked the flexibility that Stargazer offered in choosing speech commands; however, they also expressed that the uncertainty in ``\textit{what the robot actually understands}'' (P1) could make them overthink what to say.

\subsubsection{Preliminary Performance Measures}
We report objective measures for several aspects of the Stargazer prototype based on data logs and human labels of study session recordings.
Note these measures were only meant to obtain an initial understanding of the technical approaches we used.
We included recordings of the rehearsal and the final takes (if available) in the analysis.   
The total duration of the video analyzed was 71 minutes and 22 seconds.
%Future work can design more thorough evaluation conditions for 
% We manually labelled participants' gestures and utterances in study session recordings and compare the labels 

\noindent\textbf{Pointing Gestures.} We compared the human labels for participants' gestures and detection result logs. The precision for pointing detection was 91.2\%, and the recall was 86.6\%.

\noindent\textbf{Speech Commands} We calculated the metrics for speech commands, labeling on a per-sentence basis (i.e., ``you can take a close view here'' as one sentence). 
There were in total 41 sentences labeled as suggesting ``tight framing'' by humans and 40 by Stargazer. 
The precision for the label was 90.0\%, and the recall was 87.8\%. 
There were in total 18 sentences labeled as suggesting ``high-angle'' by humans and 17 by Stargazer. 
The precision for the label was 94.1\%, and the recall was 88.9\%. 
We also calculated the delay between participants finishing a sentence with a particular intention (e.g. tight framing) and the time the robot received the corresponding state change command.
The average delay was 2.4s (SD=2.3s). 
The large variance might be due to the variance in speech recognition results.

The goal of the Stargazer prototype is to demonstrate and evaluate the concept of directing camera operations by following instructor cues.
We anticipate that with the advance in machine learning and sensing technologies, the performance of the techniques we applied will continue to improve.

\section{Discussion}
Drawing on evaluation results and observations, in this section, we discuss the challenges in interpreting instructor intent for camera robots, routes to support other content creation workflows, and finally, limitations and future work. 

%\subsection{Expressiveness vs. Ease-of-Use, Automation vs. Control}
\subsection{Interpreting Instructor Intent}

One goal of our interaction designs for Stargazer is to identify and take advantage of instructor cues that clearly express their intent in camera configuration.
While our design received encouraging feedback from the evaluation, we also noted two general challenges for future camera robots that interpret instructors' or other actors' intent to assist in video creation.

The first challenge is to understand instructor intent that is not well articulated.
Although participants generally did not find incorporating camera control cues disrupted instruction, they still have to ``\textit{actively remember when to use what}'' (P3).  
P3 expected future robots to be more autonomous, ``\textit{(content creators) can just do their things and not think about there is a camera}'' (P3).
P6 suggested that future robots could introduce more variance in camera motion that fits the current scene without instructors' commands.  
A camera robot that can produce quality footage without deliberate human input should understand what is to be captured and how to capture them. However, the instructor may not want to specify them or do not know how to articulate them clearly.
Such intents can be difficult to glean from instructor cues, as cues including gestures and speech are often ambiguous when taken out of their semantic (e.g., what is the current task) and temporal (e.g., the instructor's actions so far) contexts~\cite{freeman2016do}.
One exciting opportunity for this challenge may be a data-driven approach that infers intent based on the semantic content of tutorials in the format of scripts, storyboards, or a history of the objects that the instructor has interacted with so far.

The other challenge is to build personalized robot behaviors that adapt to an instructor's preferences and style.
Content creators trying to explore the frontier may seek camera shots that are ``out-of-distribution'' rather than the most common.
For these explorers, programming-by-demonstration or other more explicit robot programming methods may help complement data-driven or heuristics-based approaches. 

\subsection{Support Other Content Creation Workflows}
Stargazer currently captures videos with a single continuous take.
\blue{While one of the participants (P6), a filmmaker, highlighted the aesthetic characteristics of one-shot videos, this is not the only format that how-to video creators adopt.
For example, P2, who runs a skateboarding social media channel, discussed the opportunity to capture skateboard drills with Stargazer and pointed out that most of his videos are heavily edited short clips for mobile consumption.
He must also perform and film the same drill several times until he gets satisfactory results.

Many content creators edit raw footage for conciseness, clarity, and visual quality.
%The instructor may choose to break the whole process of the task into segments and film them separately.
For example, some content creators might want to remove parts of the footage where the robot transitions from tracking one subject to another.
Future camera robots could assist with their workflows using the robots' data logs.
Algorithms could automatically remove these parts of the videos in post-processing based on the robots' state history.
Alternatively, the system could cut to a second, fixed camera's feed while the robot is in transition to avoid delay.}
For instructors who need multiple takes of the same process, future systems could provide interfaces for instructors to plan the robot's motion in future takes based on the recording of the first take.
% Talk about the delay and how to fix this with a second camera

\subsection{Limitations and Future Work}

\blue{The robot arm's kinematic constraints, including its fixed base and finite reach, place limits on the range of possible camera positions.
Stargazer's range of possible camera positions is sufficient for tabletop-scale activities, but not for tasks involving multiple distant locations in a room.
These constraints also restrict the use of certain camera angles (e.g. point-of-view, top-down) and camera movements (e.g. 180\degree{} orbit). 
%It also poses challenges for supporting a wide range of camera angles, such as point-of-view.
%Recent research in robot tele-operation proposed using a drone to provide a dynamic viewpoint for a robot arm performing physical tasks.
In future work, we are interested in the potential of camera drones~\cite{senft2022method} and mobile manipulators~\cite{ahn2022can} to assist with filming tasks that take place in larger environments from a wider variety of angles.}
%A promising alternative to camera drones is mobile manipulators, i.e. a robot arm on a mobile platform. 

\blue{The cues that Stargazer reacts to do not include all the rich nuances in instructors' intents. 
For example, some participants attempted to trigger object shots with deictic gestures such as giving or showing objects to the camera, which were not part of the cues that Stargazer recognizes.
We observed that most participants went through a shift of their mental models about working with Stargazer, during the training phases or the first takes. 
They began to consciously take into account what the robot could understand when delivering the tutorials, as P5 put it, ``\textit{I started to see this is a performance of me and the robot together.}''
Future research could investigate methods to detect diverse and subtle intents, possibly by combining multi-modal signals including gaze, posture, and speech.}

\blue{
%Our video analysis and interaction design for Stargazer focused on understanding and producing videos that focus on instructional content. 
Our focus in the video analysis and interaction design for Stargazer was on understanding and producing videos that emphasize instructional content.
Other types of physical-skill-focused videos, such as creative livestreams, could include significant non-instructional components like socializing and performing~\cite{fraser2019sharing} and could be several hours long. 
We plan to further study the content in these videos and their creators' camera control needs to identify additional cues and robot behaviors for Stargazer to better support non-instructional content.}

\blue{The dataset for our video analysis used the number of views as a proxy for sufficient video quality, but we did not directly assess the quality of these videos.
Thus, our data includes common instructor practices, but not necessarily the best practices. 
Datasets with comprehensive quality metrics could be constructed in the future to inform guidelines for tutorial video filming. }

As a research prototype, the current Stargazer prototype relies on an expensive, general-purpose robot arm and a suite of external sensors.
As robot-assisted filming becomes more prevalent, there could be demand for more affordable, easier-to-use, and lighter-weight specialized robots for content capture.
Future work could explore the technical and interaction design of such robots, and their implications for content creation on a broader scale.
\blue{An alternative approach is to apply a subset of Stargazer's camera control interactions to ready-to-use technologies, such as camera setups consisting of multiple fixed cameras.}

We apply a recovery motion to free the Stargazer robot from configurations close to kinematic singularities~\cite{donelan2010kinematic} or joint limits.
While the recovery motion is brief, it compromises the fluid cinematic continuity.
Computing fully singularity-free motion is still an active research area in real-time robot motion synthesis~\cite{rakita2018relaxedik}. 
A promising solution to this problem is to relax the constraints on one or more degrees of freedom in motion planning.
In the context of robot cinematography, an example of this approach is to ignore the camera-to-subject distance.
%This calls for the interesting research question on how to make the trade-off between cinematic quality and motion robustness.
%, i.e. for a particular scene setup, what are some less important camera motion constraints that can be momentarily compromised so that the camera robot can work reliably. 

While the videos captured by Stargazer were positively received by the instructors themselves, \blue{we recognize that participants could display self-evaluation bias towards their own work.}
\blue{Further evaluations could involve external experts on the skill taught and cinematography experts to counter such possible bias.
We also note that the design and evaluation of Stargazer have not yet taken the perspectives of the end consumer of how-to videos. 
Our future work will investigate learners' perceptions of robot-captured videos and compare them to baseline approaches such as videos filmed with fixed cameras.
We are also interested in interaction designs for learners to influence the behaviors of camera robots in the context of live how-to tutorials.}
%We are also interested in presenting the produced by Stargazer to professional cinematographers to learn their perspectives on the potentials and challenges for video capture with a camera robot.
\section{Conclusions}

In this paper, we proposed Stargazer, a new approach to capturing dynamic how-to video content with a camera robot that reacts to subtle cues from the instructor.
Stargazer's camera automatically follows the hands or face of the instructor, and transitions between these subjects of interest as the instructor suggests.
The instructor can use gestures and speech that are ostensibly addressed to the audience to signal to Stargazer to change the subject of interest it follows, its camera framing, and camera angle.
Our interaction design for Stargazer was informed by an analysis of 50 online how-to videos.
A user study with six participants, including two professional filmmakers, showed that physical skill instructors can successfully create how-to video content with Stargazer and were satisfied with the produced videos' quality.
They also did not find the interaction disruptive to instruction delivery.
Stargazer explores the potential of robot-assisted video filming for how-to videos, and we hope it can inspire future research in human-robot collaborative content creation.

% With Stargazer, an instructor can create compelling videos... 
%Our approach does techincal stuff and list of features
% we demonstrate these things in the paper
% did an evaluation
% results

\begin{acks}
Authors would like to thank the support from the participants, the anonymous reviewers, and NSERC grants RGPIN-2017-06415, DGDND-2017-00093, IRCPJ 545100-18, SMFSU 549149-2020, and RGPIN-2017-04883.
\end{acks}

\bibliographystyle{ACM-Reference-Format}
\bibliography{main}

\appendix
\section{Additional Details on Robot Motion Planning}
We provide more details on motion planning, in addition to the content in ~\ref{sec:motion-planning}, including the calculation of the orientation cost, and the coordinate system we used in optimization.

\subsection{Orientation Cost} \label{sec:appendix-orientation}
As introduced in~\ref{sec:cost-function}, for the orientation cost term we calculate it based on the different between the orientation of the next camera position, $\mathbf{v_{t+1}}$ and the target camera orientation $\mathbf{v_o}$.
Both vectors are in the ground plane (x-y plane in Figure~\ref{fig:appendix-alg}). 
We describe the procedure to compute these two vectors for the case that the camera should look perpendicular at the link between the instructor's two shoulder joints ($v_{sh}$ indicates its direction). 
The computation for truck movements, where the camera orientation should be perpendicular to the finger trace, is similar.

To compute $\mathbf{v_o}$, we first get one particular vector, $v_{ba}=v_{sh} \times v_g$, that is perpendicular to $v_{sh}$.
$v_g=(0, 0, -1)$ is the gravity direction.
Then $v_o$ is the projection of $v_{ba}$ onto the ground plane, normalized.
To compute $v_{t+1}$, we simply project the camera orientation $(x_{s}-x_{t+1})/\|x_{s}-x_{t+1}\|$ onto the ground plane. 
Then the orientation cost is square of the norm of the difference between these two vectors. 
That is $c_o=\| \mathbf{v_{t+1}} - \mathbf{v_o} \|^2$

\subsection{Polar Coordinate System Used for Constrained Optimization} \label{sec:appendix-polar}
\begin{figure}[t]
  \centering
  \includegraphics[width=\columnwidth]{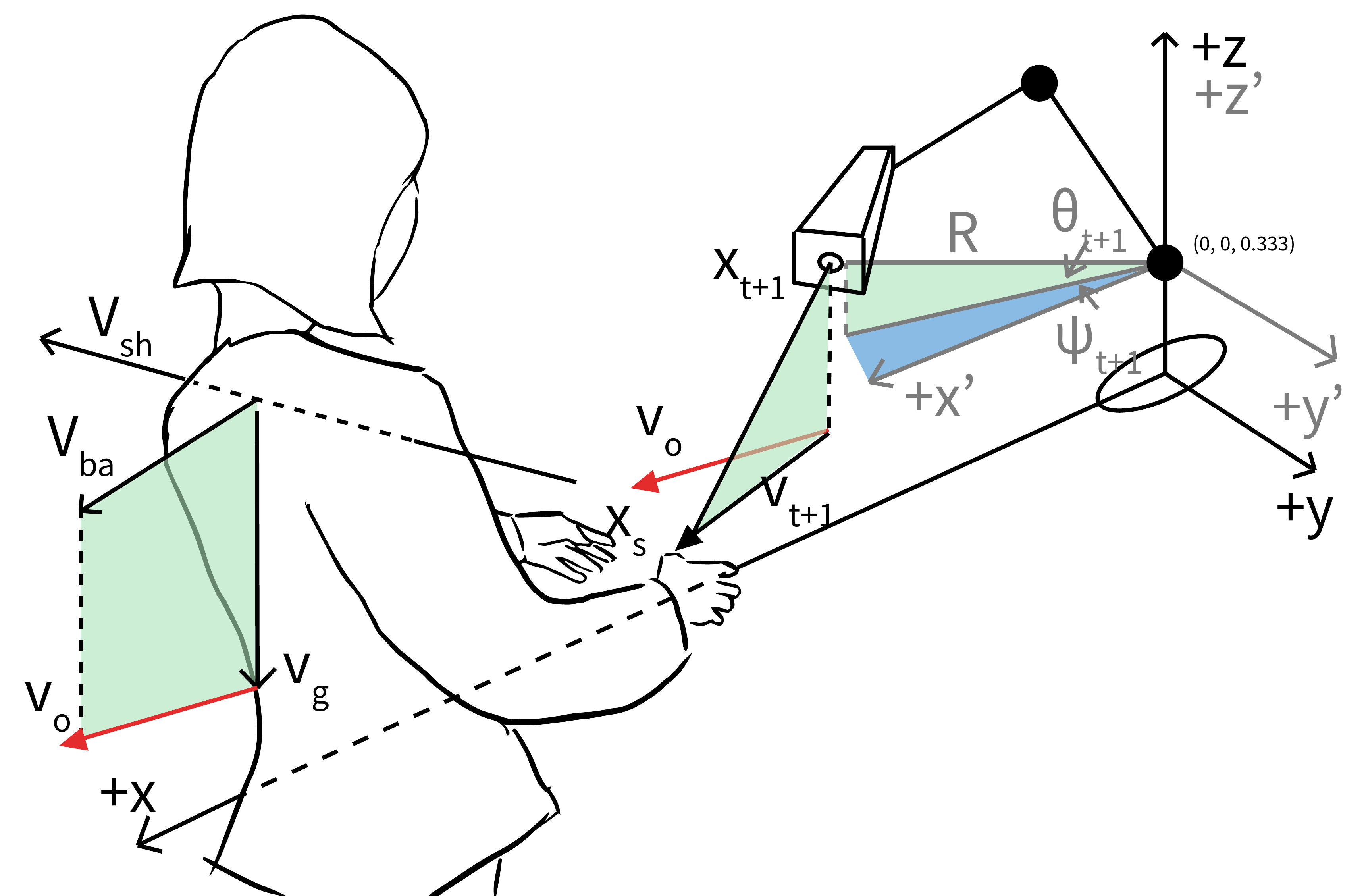}
  \caption{ The computation of the desired camera direction $v_o$ (in red) and the polar coordinate system (axes in grey) we used for optimization. Angles in vertical planes are in green, angles in horizontal planes are in blue.}
  \Description{This figure shows an illustration about the computation of the desired camera direction and the polar coordinate system we used for
optimization.}
  \label{fig:appendix-alg}
\end{figure}

To simply the constraint condition in the optimization, we convert the next position of the camera from Cartesian coordinates to polar coordinates.
The origin of our polar coordinate system is at the second joint of the robot (Figure~\ref{fig:appendix-alg}).
The Cartesian coordinates of this point in the robot frame is (0, 0, 0.333m), according to the specification of the Panda robot. 
The x, y, z axis of the polar coordinate system is parallel to the robot coordinate system.
$R$ is the distance between the camera and the origin.
$\theta$ is the angle that is on the vertical plane.
$\psi$ is the angle that is one the horizontal plane.
We chose $R \in (0.36m, 0.66m)$, $\theta \in (-\pi/10, 0.4\pi)$, and $\psi \in (-0.4\pi, 0.4\pi)$ as the bounds in optimization.

\end{document}